\documentclass[traditabstract,10pt]{aa}

\usepackage[T1]{fontenc}
\usepackage{amsmath, amssymb, amsfonts, latexsym}
\usepackage[breaklinks,colorlinks,citecolor=blue]{hyperref} 
\usepackage{ifthen}
\usepackage[dvipsnames,table]{xcolor}
\usepackage[varg]{txfonts}
\let\vec\mathbf
\usepackage{ulem}

\usepackage{graphicx}
\graphicspath{{Figures/}}
\usepackage[skip=5pt]{caption}

\usepackage[nonamebreak]{natbib}
\bibpunct{(}{)}{;}{a}{}{,}
\providecommand{\sorthelp}[1]{}

\newcommand{\plik}{{Plik}}
\newcommand{\hillipop}{{HiLLiPoP}}
\newcommand{\lollipop}{{LoLLiPoP}}

\newcommand{\commander}{{Commander}}
\newcommand{\camspec}{{Camspec}}
\newcommand{\PLK}{\textit{Planck}}
\newcommand{\SPT}{{SPT}}
\newcommand{\SPTg}{{SPT-3G}}
\newcommand{\ACT}{{ACT}}
\newcommand{\PACT}{{P-ACT}}
\newcommand{\lcdm}{$\Lambda$CDM}

\def\Alens{{\ifmmode{A_\mathrm{L}}\else{$A_\mathrm{L}$}\fi}}
\def\Mnu{{\ifmmode{\textstyle \Sigma m_\nu}\else{$\textstyle \Sigma m_\nu$}\fi}}
\def\Ok{{\ifmmode{\Omega_{K}}\else{$\Omega_{K}$}\fi}}
\def\Neff{{\ifmmode{N_\mathrm{eff}}\else{$N_\mathrm{eff}$}\fi}}
\def\GHz{\ifmmode $\,GHz$\else \,GHz\fi}
\def\muK{\ifmmode \,\mu$K$\else \,$\mu$\hbox{K}\fi}

\defcitealias{tristram:2024}{T24}
\defcitealias{Louis:2025}{L25}
\defcitealias{Camphuis:2025}{C25}
\defcitealias{Beringue:2025}{B25}

\begin{document}

\title{Combining cosmic microwave background datasets with consistent foreground modelling}
\titlerunning{Combining CMB datasets with consistent foreground modelling}

\author{
M.~Tristram\inst{1}
\and M.~Douspis\inst{2}
\and A.~Gorce\inst{2}
\and S.~Henrot-Versill{\'e}\inst{1}
\and L.~T.~Hergt\inst{1}
\and S.~Ilic\inst{1}
\and L.~McBride\inst{2} 
\and M.~Mu{\~n}oz-Echeverr{\'i}a\inst{3}
\and E.~Pointecouteau\inst{3}
\and L.~Salvati\inst{2}
}

\institute{
Universit{\'e} Paris-Saclay, CNRS/IN2P3, IJCLab, 91405 Orsay, France
\and
Universit{\'e} Paris-Saclay, CNRS, Institut d'Astrophysique Spatiale, 91405 Orsay, France
\and
IRAP, CNRS, Universit{\'e} de Toulouse, CNES, UT3-UPS, Toulouse, France
}

\abstract{
We present a joint cosmological analysis combining data from the \PLK\ satellite, the Atacama Cosmology Telescope, and the South Pole Telescope. We construct a unified likelihood that reproduces the measured temperature and polarisation power spectra by jointly modelling the cosmic microwave background (CMB) signal, Galactic and extragalactic foregrounds, and instrumental systematics across all datasets. We reduce reliance on external priors by combining datasets and improve the robustness of parameter estimation by marginalising over the choice of foreground templates. Within this joint analysis, \lcdm\ parameters exhibit remarkable stability with respect to variations in foreground modelling. Parameters for cosmological extensions are more sensitive to these assumptions, with uncertainties increased by up to 35\% in the neutrino sector after marginalising over foreground models. In contrast, the determination of foreground parameters depends more strongly on the assumptions made about the underlying foreground models. Overall, this work demonstrates the feasibility and reliability of a fully joint analysis of current CMB experiments and emphasises the importance of consistent and accurate foreground modelling for the scientific goals of next-generation, high-sensitivity CMB surveys.
}


\date{\today}

\maketitle

\section{Introduction}

The cosmic microwave background (CMB) remains one of the most powerful probes of the early Universe, providing precise measurements of the fundamental parameters that govern cosmology. Over the past decade, multiple ground-based and space-borne experiments -- most notably the \PLK\ satellite, the Atacama Cosmology Telescope (\ACT), and the South Pole Telescope (\SPT) -- have independently mapped the CMB sky with increasing sensitivity, resolution, and frequency coverage. 

\begin{figure}[htbp!]
	\centering
	\includegraphics[width=.9\columnwidth]{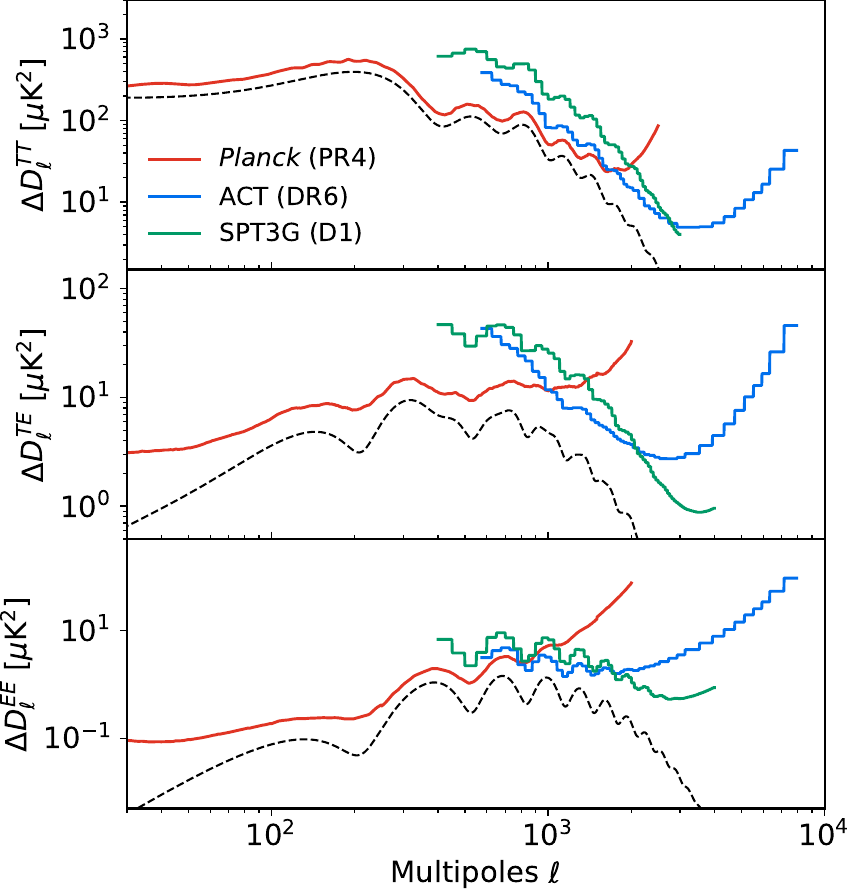}
	\caption{Uncertainties on angular power spectra $TT$, $TE$, and $EE$ for the \PLK\ PR4 \citep[red;][]{tristram:2024}, \ACT\ DR6 \citep[blue;][]{Louis:2025}, and \SPTg\ D1 \citep[green;][]{Camphuis:2025} datasets compared to full-sky cosmic variance (dashed line). Uncertainties are averaged from all cross-frequency spectra according to their respective covariance matrix.}
	\label{fig:sigl}
\end{figure}

Each of these experiments contributes complementary strengths: \PLK\ offers full-sky coverage and excellent sensitivity at large angular scales; \ACT\ and \SPT\ provide high-resolution measurements on smaller scales, with deeper observations over limited regions of the sky. 
Fig.~\ref{fig:sigl}  illustrates the relative sensitivity of each experiment across angular scales. We show the uncertainties on the $TT$, $TE$, and $EE$ power spectra as a function of multipole for \PLK\ \citep[PR4,][]{tristram:2024}, \ACT\ \citep[DR6,][]{Louis:2025}, and \SPTg\ \citep[D1,][]{Camphuis:2025}. This comparison highlights the complementarity of the datasets: \PLK\ dominates at low multipoles, while \ACT\ and \SPT\ significantly improve constraints at smaller angular scales, particularly in polarisation.

However, these differences in sky coverage, angular resolution, and frequency channels introduce critical challenges to a consistent joint analysis.
Although each collaboration has developed its own foreground model, residual foregrounds (Galactic dust, thermal and kinetic Sunyaev-Zel'dovich effects, the cosmic infrared background, and extragalactic sources) are common to all experiments. Modelling these shared signals differently across likelihoods makes the resulting joint constraints difficult to interpret and reduces their reliability. \citet[][hereafter \citetalias{Beringue:2025}]{Beringue:2025} explore aspects of this problem by applying the \ACT\ foreground model to \PLK\ and \SPT\ data.

In this work, we present a coherent combination of the \PLK, \ACT, and \SPT\ datasets, jointly analysing their temperature and polarisation power spectra within a common model for both cosmological and foreground contributions. 
This approach takes advantage of the full constraining power of each experiment, taking into account differences in frequency coverage and resolution. By fitting a unified model to all three datasets simultaneously, we minimise reliance on external priors, tighten constraints on the parameters of the cosmological-constant-dominated cold dark matter (\lcdm) model, and, for the first time, provide uncertainties associated with foreground modelling and study their impact on both \lcdm\ and selected extended parameters.

This paper is organised as follows. We describe the likelihood construction in Sect.~\ref{sec:lik} and the datasets in Sect.~\ref{sec:datasets}. We detail the instrumental effects in Sect.~\ref{sec:nui}  and the sky model in Sect.~\ref{sec:model}. We present the results of the combined analysis for \lcdm\ parameters in Sect.~\ref{sec:lcdm} and in Sect.~\ref{sec:foregrounds}, we discuss the impact of the foreground modelling. We derive constraints on extensions to \lcdm\ in Sect.~\ref{sec:extensions}. Finally, we conclude in Sect.~\ref{sec:conclusion}.

\section{Likelihood}
\label{sec:lik}
Information at large angular scales ($\ell<30$) comes from CMB maps obtained after component separation of the \PLK\ data, and we treated it separately from the smaller scales.

At high multipoles ($\ell \gtrsim 30$), current CMB likelihoods are based on frequency cross-spectra, and we adopted a Gaussian approximation, as in the original implementations of the \PLK, \SPT, and \ACT\ likelihoods.
Although the exact distribution of auto- and cross-spectra is non-Gaussian \citep[scaled-$\chi^2$ and more complex forms; see][]{mangilli:2015}, the large number of modes at high-$\ell$ makes the Gaussian approximation sufficiently accurate \citep{carron:2013}.
However, cutting the sky to reduce Galactic contamination induces correlations between the $C_\ell$ values, which necessitates the estimation of a full covariance matrix. The cross-spectra from the frequency maps $\vec{C}_\ell$ are then compared with the data model $\vec{\hat C}_\ell$, in the likelihood
\begin{equation}
    -2 \ln \mathcal{L} = (\vec{C} - \vec{\hat C})^T \, \tens{\Sigma}^{-1} \, 
                         (\vec{C} - \vec{\hat C})
    \label{eq:likelihood}.
\end{equation}
The covariance matrix $\Sigma$ includes both instrumental and cosmic variance and is, in principle, model-dependent. 
We include correlations between multipoles and cross-frequency spectra in the likelihood. However,  we did not consider any correlation between the datasets. 
As a consequence, to avoid overlap in multipoles, we restricted the multipole range of the \ACT\ spectra when combined with \PLK. 
Specifically, we used \PLK\ data for multipoles from $\ell = 2$ to 2000 for $TT$, 1500 for $TE$, and 1000 for $EE$, while we used ACT data for multipoles above these thresholds.
We neglected the correlations between \PLK\ and \SPT, as the two experiments overlap over less than 5\% of the sky.

We analysed each dataset both independently and in combination. The model includes the CMB signal, together with residuals of diffuse and unresolved foreground emissions in the power spectra and instrumental systematic effects.

\section{Datasets}
\label{sec:datasets}

\subsection{Planck}

We used the PR4 maps as the \PLK\ sky measurements in this analysis. The PR4 maps were produced with the {\tt NPIPE} processing pipeline, which creates calibrated frequency maps in temperature and polarisation from the \PLK\ Low Frequency Instrument (LFI) and High Frequency Instrument (HFI) data. As described in \citet{planck2020-LVII}, {\tt NPIPE} processing includes several improvements, resulting in lower levels of noise and systematics in both frequency maps and component-separated maps at essentially all angular scales, as well as notable improvements in internal consistency between the various frequencies.

To cover the very low multipoles ($\ell < 30$), we used the \commander\ $TT$ likelihood~\citep{planck2016-l04} and, for polarisation, we added the low-$\ell$ likelihood \lollipop\ introduced in \cite{tristram:2021,tristram:2022}.

At high multipoles, the data consist of 15 cross-spectra computed from the two detector-set (detset) maps at three frequencies (100, 143, and 217\,GHz). We further combined these into six cross-frequency spectra by applying a calibration associated with each detset map. The resulting \PLK\ likelihood is a binned version of the \PLK\ likelihood, \hillipop\, presented in \citet{tristram:2024} \citepalias[hereafter][]{tristram:2024}.

\subsection{Atacama Cosmology Telescope}
The DR6 data release is based on the 90, 150, and 220 GHz data from the 2017--2022 observing seasons of the AdvancedACT camera. The survey consists of 19\,000 square degrees of the sky, with a median combined depth of 10\,{$\mu$K$\cdot$arcmin}. We refer to \citet{Naess:2025} for a description of the frequency maps, as well as the data reduction pipeline.

In this study, we adopt a likelihood implementation closely aligned with the \texttt{MFLike} software developed for the Simons Observatory and used in \citet{Louis:2025} (hereafter \citetalias{Louis:2025}), retaining the same set of instrumental parameters. 
The data include five different sets of maps from three different wafers: pa5 (90 and 150\GHz), pa6 (90 and 150\GHz), and pa4 (at 220\GHz), the latter being considered only in temperature. The main modification concerns the foreground treatment, which we adapted to allow unified modelling across all datasets.
We verified that under identical model assumptions and priors, our implementation reproduces the results of \citetalias{Louis:2025} up to machine precision.

\subsection{South Pole Telescope}
The \SPT\ data used in this analysis correspond to the \SPTg\ D1 dataset, which includes 2019 and 2020 observations of the \SPTg\ Main field and covers the 95, 150, and 220\,GHz frequency bands. The data released include $TT$ band powers that span the angular-multipole range $400 < \ell < 3000$ and $TE$/$EE$ band powers that cover $400 < \ell < 4000$. The cross-frequency spectra are binned into band powers of width $\Delta\ell = 50$ \citep[][hereafter \citetalias{Camphuis:2025}]{Camphuis:2025}. 
We validated the likelihood implementation used in this work against the official \texttt{candl} likelihood up to machine precision.

\section{Instrumental effects}
\label{sec:nui}

In the combined \PLK, \SPT, and \ACT\ likelihood, we implemented the same systematic corrections as used in the individual likelihoods. These corrections account for a variety of well-understood instrumental effects:

- Calibration.
For each dataset, we sampled inter-calibration coefficients relative to a reference map: the first 143\GHz\ detset for Planck, the 150\GHz\ channel for \SPT, and  pa5 at 150\GHz\ for \ACT. We varied coefficients for the remaining maps (five detsets for \PLK; 90 and 220\GHz\ for \SPT; five wafers for \ACT), using Gaussian priors from the original analyses (Table~\ref{tab:nuipar}).
We also included absolute calibration parameters, with Planck as the reference: we constrained $A_{\PLK}$ with a strong prior from the orbital dipole uncertainty, while we determined the \ACT\ and \SPT\ calibrations relative to it.

- Polarisation efficiency.
We adjusted the polarisation efficiencies of each map by a multiplicative factor $\rho$, which we sampled independently with flat priors.

- Super-sample lensing amplitude.
We parametrised the super-sample lensing amplitude in \SPT\ by an additional parameter $\kappa$ constrained by a Gaussian prior. 
For \ACT\ and \PLK, the bias on the angular power spectra is negligible. However, \ACT\ includes a second-order correction in the $C_\ell$ covariance matrix to account for lensing super-sample covariance arising from fluctuations in the mean convergence across the observed area.

- Aberration.
For SPT, we corrected the aberration in the likelihood with a fixed parameter, while for \ACT\ the power spectra are already corrected using simulations. In the case of \PLK, the aberration effect is negligible.

- Bandpass integration.
To compute the foreground amplitudes for a given spectral energy density (SED), we integrated over the bandpass of each channel. This is commonly approximated by assigning an effective frequency to each component and channel, as in the \SPT\ and \PLK\ likelihoods. However, \citetalias{Louis:2025} show that bandpass uncertainties in ACT-DR6 contribute significantly to the overall error budget in the recovery of foreground parameters. To account for this, we sampled six additional bandpass shift parameters, $\Delta_{\rm bp}$, constrained by strong priors derived from instrumental calibration. In addition, the \ACT\ team incorporates chromatic beam window functions to model the frequency dependence of the beam, an approach we adopted following \citet{Giardiello:2025}.

- Beam uncertainties.
Beam uncertainties are negligible for the \PLK\ dataset \citep[see][]{planck2013-p08}. For \ACT, they are included as an additional contribution to the covariance matrix \citepalias{Louis:2025}. In the case of \SPT, beam uncertainties are propagated by fitting the amplitude $\beta_i$ of the first nine eigenmodes of the beam covariance matrix. Furthermore, the \SPT\ likelihood includes additional freedom to rescale the polarisation amplitude of the sidelobes at each frequency \citepalias{Camphuis:2025}.

- Temperature-to-polarisation leakage.
For \PLK, the spectra are already corrected and the associated uncertainty is negligible. 
For \ACT,  a correction is included in the covariance matrix, while the \SPT\ likelihood propagates the uncertainty by fitting a contribution from quadrupolar beam leakage, modelled analytically for each frequency \citepalias{Camphuis:2025}.
\\

Table~\ref{tab:nuipar} lists the nuisance parameters and their associated priors. When combining all datasets, we sampled a total of 48 nuisance parameters. Although the large number of parameters can pose challenges for sampling the likelihood, particularly due to degeneracies and long correlation lengths, they are essential for accurately propagating instrumental uncertainties that affect the angular power spectra. Including these parameters ensures that the cosmological constraints properly account for residual systematics and modelling uncertainties across all experiments.

\section{Sky model}
\label{sec:model}

We now present the sky model ($\vec{\hat C}_\ell$) used in the CMB likelihood (Eq.~\ref{eq:likelihood}). Residual diffuse and unresolved components are modelled directly in the likelihood, with contributions in temperature ($T$) and polarisation ($P$) treated as follows: 
\vspace{-0.2cm}
\begin{itemize}
    \item Galactic emissions ($T$, $P$),
    \item the cosmic infrared background CIB ($T$),
    \item the Sunyaev-Zel'dovich effect: thermal tSZ, kinetic kSZ, and correlation with infrared galaxies tSZ$\times$CIB ($T$),
    \item radio sources ($T$, $P$).
\end{itemize}
\vspace{-0.2cm}
The cosmic infrared background and the Sunyaev--Zel'dovich effects as well as their correlation are assumed to be unpolarised.
Our model does not include extragalactic carbon monoxide (CO) \citep{maniyar:2023,Kokron:2024}, as its amplitude is expected to be subdominant to SZ$\times$CIB and kSZ and its modelling remains uncertain, particularly regarding its distinction from other extragalactic components such as the CIB. 

In the original implementations of the CMB likelihoods, extragalactic foregrounds are typically modelled using fixed angular power spectrum templates, whose amplitudes are fitted simultaneously with the CMB signal. These templates are often derived from external observations (e.g. SZ or CIB measurements), computed from simulations, or based on simplified assumptions (\mbox{power-law}).

In this study, we ensured that the foregrounds were consistently modelled for all datasets. Foreground templates are shared by all CMB likelihoods, and extragalactic emissions (this includes the CIB, tSZ, kSZ and tSZxCIB) are fitted consistently using the same parameters across the datasets. Table~\ref{tab:fgpar} provides a summary of all parameters related to the foreground modelling.

Crucially, combining multiple experiments with complementary frequency coverage, angular resolution, and noise properties significantly enhances our ability to disentangle foreground contributions. \PLK\ is particularly effective for Galactic components and the CIB, while the higher resolution and small-scale sensitivity of \ACT\ and \SPT\, allow tighter constraints on the tSZ, kSZ, and point-source signals. 
This complementarity breaks degeneracies that persist in single-dataset analyses, allowing many of the informative priors on foreground parameters to be relaxed or removed (see Table~\ref{tab:fgpar}), resulting in a data-driven sky model (see Fig.~\ref{fig:par_extragal}).

\subsection{Cosmic microwave background}
We computed the CMB signal, $\hat{C}^{\rm CMB}_\ell$ by numerically solving the background and perturbation equations for a given cosmological model. We considered a \lcdm\ model with six free parameters: the present-day baryon density ($\Omega_bh^2$), cold dark matter density ($\Omega_\mathrm{c}h^2$), the angular size of the sound horizon at recombination ($\theta_{\rm s}$), the reionisation optical depth ($\tau$), and the amplitude ($A_{\rm s}$) and the spectral index ($n_{\rm s}$) of the primordial scalar power spectrum. We extend the model with single-parameter extensions in Sect.~\ref{sec:extensions}.

We used the Boltzmann solver \texttt{CAMB} \citep{Lewis:2000} and the \texttt{CosmoPower} emulator \citep{Mancini:2022} to compute the CMB power spectra. We primarily used the emulator to accelerate the initial MCMC explorations, while we validated the final results with the full Boltzmann codes to ensure accuracy.
To meet the precision requirements of high-resolution surveys, we adopted the accuracy settings of the ACT-DR6 analysis \citepalias{Louis:2025}.

Unlike the original analyses, we did not impose any prior on the reionisation optical depth, $\tau$, even when considering datasets covering only high multipoles (\ACT\ or \SPT) individually.

\subsection{Galactic emissions}
In modelling Galactic emission, CMB likelihoods include only Galactic dust and neglect synchrotron or free-free emission in the frequency range of our datasets \citep{planck2016-l04}.

At first order, the frequency dependence of Galactic dust grains can be modelled as a greybody, $a_\nu^{\rm dust} = \nu^{\beta_{\rm d}} B_\nu(T_{\rm d})$. 
Throughout the frequency range of our datasets, the Rayleigh–Jeans approximation applies, leading to a strong degeneracy between $\beta_{\rm d}$ and $T_{\rm d}$. Consequently, the greybody temperature is commonly fixed to $T_{\rm d}=19.6$\,K.
The spectral indices can be fixed or marginalised over with Gaussian priors.

The power spectrum of the Galactic dust emission was measured from \PLK\ full-sky maps at high-frequency channels (353, 545, and 857\,GHz) for temperature and polarisation \citep{planck2014-XXX,planck2016-l11A}. 
The dust spectrum is well described by a red index power law, $\alpha_{\rm d}$, which means that the dominant signal is at low multipoles.
In the \ACT\ likelihood~\citepalias{Louis:2025}, the slope of the power law is fixed to $\alpha_{\rm d}^{TT} = -2.6$ and $\alpha_{\rm d}^{TE/EE} = -2.4$. For \SPT, \citetalias{Camphuis:2025} marginalise over the index with strong priors, $\alpha_{\rm d}^{TT}=\mathcal{N}(-2.53,0.05)$ and $\alpha_{\rm d}^{TE/EE}=\mathcal{N}(-2.42,0.04)$, based on an analysis of \PLK\ data over the \SPTg\ survey patch. 
For the \PLK\ likelihoods, the masking procedure is adapted to each frequency, so the dust model can involve three different indices. In \citetalias{tristram:2024}, for PR4, the indices are $-2.65$, $-2.55$, $-2.45$ for sky fractions of 80, 70, and 60\%, respectively. In \citet{planck2016-l05}, the PR3 likelihood used an ad hoc effective analytical model, $C_\ell \propto (1+h\ell^k \exp{-\ell/t}) \times (\ell/\ell_p)^{\alpha_{\rm d}}$, with free parameters $h$, $k$, $t$, and $\alpha_{\rm d}$ fitted to \PLK\ 545\GHz\ data and adapted to each mask. The recovered indices are $-2.65$, $-2.57$, and $-2.55$ at 70\%, 60\% and 50\% sky fractions, respectively. For $TE/EE$, the index is fixed to $\alpha_{\rm d} = -2.4$.

Although the index can vary slightly across the sky, dust emissions are well described by a global $\alpha_{\rm d} = -2.6$ in temperature and $-2.4$ in polarisation \citep{planck2016-l05,planck2016-l11A}. In this study, we used a common power-law spectrum to describe the dust power spectra,
\begin{equation}
	\hat{C}_\ell^{\rm dust}(\nu\times\nu') = A^{\rm dust} \, \frac{a^{\rm dust}_{\nu}}{a^{\rm dust}_{\nu_0}} \frac{a^{\rm dust}_{\nu'}}{a^{\rm dust}_{\nu_0}} \,  \left(\frac{\ell}{\ell_0}\right)^{\alpha_{\rm d}},
\end{equation}
where $\alpha_{\rm d}$ is free in temperature and fixed to $\alpha_{\rm d} = -2.4$ for $TE/EE$.
The frequency dependence follows a greybody spectrum with priors on $\beta_{\rm d}$ from \PLK\ measurements, $\beta_{\rm d}^{TT} = \mathcal{N}(1.51, 0.01)$ and $\beta_{\rm d}^{TE/EE} = \mathcal{N}(1.59, 0.02)$, and a fixed temperature $T_{\rm dust} = 19.6$\,K \citep{planck2014-XXII}.
We adapted the amplitude $A^{\rm dust}$, expressed at the reference frequency $\nu_0=150\GHz$ and the multipole $\ell_0=3000$, to each sky survey (including the three surveys for \PLK) and each polarisation mode ($TT$, $TE$, $EE$). 
In the high-multipole CMB datasets (\ACT\ and \SPT) the residual Galactic dust contribution is subdominant and therefore we imposed informative priors on $A^{\rm dust}$. Conversely, in the case of \PLK, where Galactic dust dominates the low multipoles across all frequencies, we adopted flat priors on the corresponding amplitudes.

\subsection{Cosmic infrared background}
The CIB consists of emission from unresolved dusty star-forming galaxies \citep[DSFGs;][]{puget:1996,gispert:2000,lagache:2005}. The angular power spectrum from these galaxies can be modelled as the sum of a Poisson noise term and a clustered component related to the formation of large-scale structures.

The frequency dependence of the CIB can be described as a greybody emission $a_\nu^{\rm CIB} = \nu^{\beta_{\rm CIB}} B_\nu(T_{\rm CIB})$. However, the temperature adopted in the likelihoods of different CMB datasets varies, from $T_{\rm CIB}=9.7$\,K in \ACT\ \citep[from][]{addison:2012a} to $T_{\rm CIB}=25$\,K for \PLK\ \citep[from][]{planck2013-pip56}. In its last data release, \SPT\ does not model the CIB emission law explicitly, instead using one amplitude for each cross-spectrum including frequencies above 150\GHz, while neglecting the CIB for 90$\times$90, 90$\times$150, and 90$\times$220 cross-spectra.

The angular power spectrum of the clustered CIB was initially described with a power law $\ell^{\alpha_{\rm CIB}}$ (in $C_\ell$). Here again, the index $\alpha_{\rm CIB}$ varies from $-1.25$ \citep{addison:2012a} to $-1.47$ \citep{mak2017}. 
Since 2015, \PLK\ likelihoods \citep{planck2014-a15,planck2016-l06,tristram:2024} and the recent \ACT\ likelihood \citepalias{Louis:2025}, have used a more complex model, including 1-halo and 2-halo terms, which provides an accurate description of the \PLK\ and IRAS CIB spectra from 217\GHz to 3000\GHz\ \citep{planck2013-pip56}. \citet{lenz:2019} later updated this model.
By contrast, for \SPT, \citetalias{Camphuis:2025} continue to use a power law with $\alpha_{\rm CIB}=-1.47$ from~\citet{mak2017}.
Several codes now exist to compute the angular power spectrum from a halo model \citep[see for instance][]{viero:2013,maniyar:2021,Zagatti:2024}.

\begin{figure}[!ht]
	\centering
	\includegraphics[width=.85\columnwidth]{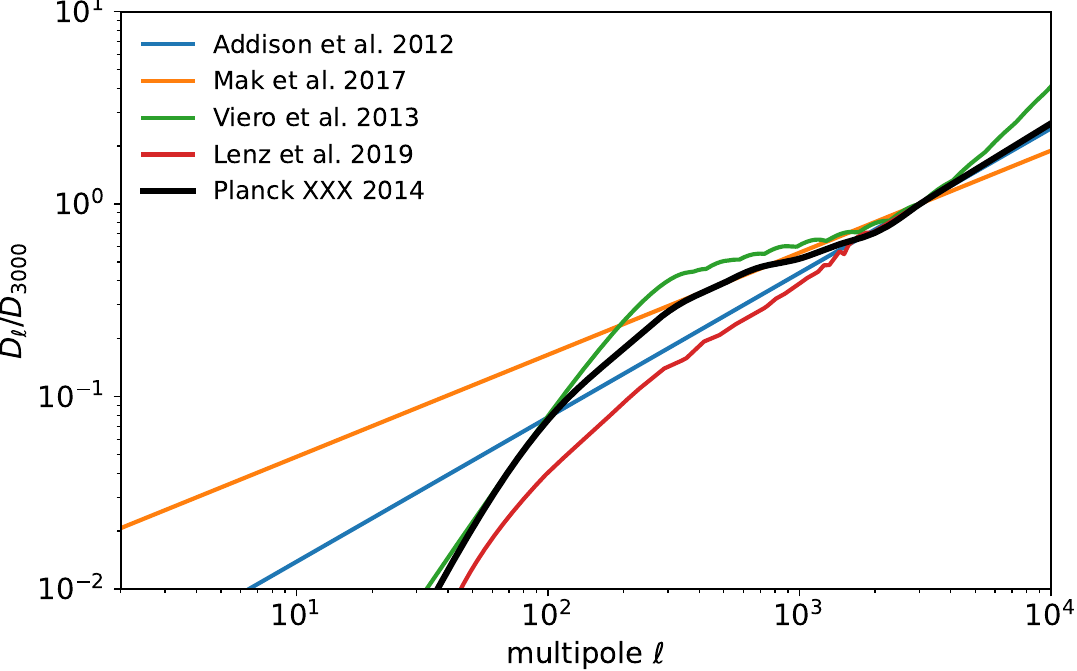}
	\caption{Templates of the CIB power spectrum normalised at $\ell_0=3000$ (the black line corresponds to the baseline template).}
	\label{fig:dl_cib}
\end{figure}

We adopted the power spectrum from \citet{planck2013-pip56} as a baseline and examined the impact of alternative $\mathcal{C}_\ell^{\rm CIB}$ shapes (Fig.~\ref{fig:dl_cib}). In the likelihoods, we rescaled the $\mathcal{C}_\ell^{\rm CIB}$ template by the amplitude of the clustered component, $A^{\rm CIB}$, and added the Poisson term, $A^{\rm IR}$, both expressed at the reference frequency $\nu_0=150$\GHz\ and multipole $\ell_0=3000$, such that
\begin{equation}
\hat{C}_\ell^{\rm CIB}(\nu\times\nu') = \frac{a_{\nu}^{\rm CIB}}{a_{\nu_0}^{\rm CIB}}\frac{a_{\nu'}^{\rm CIB}}{a_{\nu_0}^{\rm CIB}} (A^{\rm CIB} \mathcal{C}_\ell^{\rm CIB} + A^{\rm IR}).
\end{equation}
We assumed perfect correlation of the emission across the frequency range considered (95--220\GHz) and adopted a flat prior on $\beta_{\rm CIB}$ with $T_{\rm CIB}=25$\,K, except for analyses of individual datasets, where we used a Gaussian prior based on \PLK\ measurements, $\beta_{\rm CIB} = \mathcal{N}(1.75,0.06)$ \citep{planck2013-pip56}.

\subsection{Thermal Sunyaev--Zeldovich}
The tSZ effect is a spectral distortion of the CMB caused by inverse Compton scattering of photons off the hot gas in massive halos, groups, and clusters of galaxies \citep{Sunyaev:1972}. After masking the resolved clusters in the maps, the CMB likelihoods include a component accounting for residual unresolved tSZ emission.

When the relativistic correction is neglected, the frequency dependence of the tSZ effect is given by $a_\nu^{\rm tSZ} = x(e^x+1)/(e^x-1) - 4$ with $x = h\nu/(k_{\rm B} T_{\rm CMB})$.

The \PLK\ PR3 likelihoods adopt an empirical template for the tSZ power spectrum, as described in \cite{efstathiou:2012}. The \ACT\ team uses models from \citet{battaglia:2012}, while the \SPT\ team initially used a template from \citet{shaw:2010} and later adopted one derived from the Agora simulations \citep{agora}.
The \hillipop\ \PLK\ PR4 likelihood used the tSZ power spectrum modelled using a halo-based approach described in \citet{Douspis:2022} and measured by \PLK\ in \citet{planck2014-a28} and \citet{tanimura:2022}.
In addition, \citetalias{Louis:2025} propose the introduction of additional flexibility by allowing a scale dependence in the tSZ template, parametrised as $\mathcal{C}_\ell^\mathrm{tSZ} \rightarrow \ell^{\alpha_\mathrm{tSZ}}\,\mathcal{C}_\ell^\mathrm{tSZ}$.

\begin{figure}[!ht]
	\centering
	\includegraphics[width=.85\columnwidth]{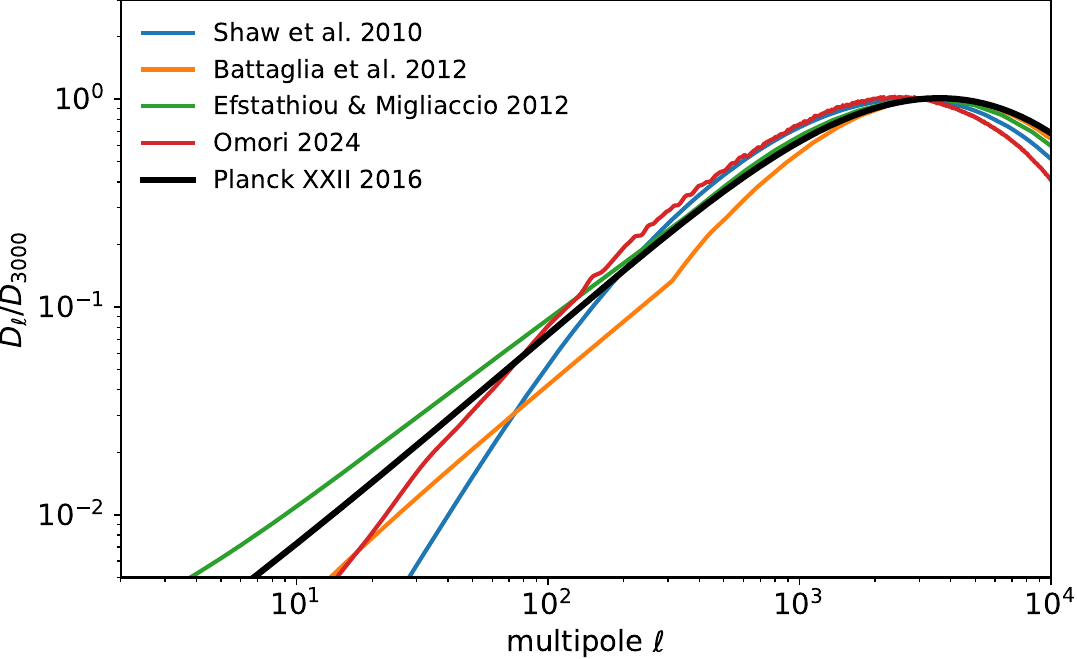}
	\caption{Templates of the tSZ power spectrum normalised at $\ell_0=3000$ (the black line corresponds to the baseline template).}
	\label{fig:dl_tsz}
\end{figure}

We adopted the power spectrum from \citet{planck2014-a28} as a baseline and assessed the impact of alternative $\mathcal{C}_\ell^{\rm tSZ}$ shapes (Fig.~\ref{fig:dl_tsz}). The model is parametrised by a single amplitude $A^{\rm tSZ}$, corresponding to the tSZ emission at the reference frequency ($\nu_0=150$\GHz) and multipole $\ell=3000$, such that
\begin{equation}
	\hat{C}_\ell^{\rm tSZ}(\nu\times\nu') = A^{\rm tSZ} \, a_{\nu}^{\rm tSZ}a_{\nu'}^{\rm tSZ} \, \mathcal{C}_\ell^{\rm tSZ}.
\end{equation}

\subsection{Kinetic Sunyaev--Zeldovich}
Due to the motion of galaxy clusters with respect to the CMB rest frame, an additional spectral distortion arises from the Doppler effect of their bulk velocity on the scattered CMB photons. This distortion is known as the kSZ effect \citep{zeldovich:1969,sunyaev:1980}, with amplitude proportional to the line-of-sight velocity $v_c$, $\Delta T_{\rm kSZ}/T_{\rm CMB} = - \tau_e v_c/c$, where $\tau_e$ is the optical depth of the cluster.
The kSZ signal shares the same blackbody spectrum as the primary CMB, making it indistinguishable in frequency space and therefore requires template-based modelling.

The kSZ power spectrum is typically computed from simulations or described using halo models. It is often decomposed into (i) a late-time, homogeneous component from scattering by free electrons in the ionised IGM after reionisation and (ii) a patchy component from scattering off ionised bubbles during reionisation.

The \ACT\ likelihood uses a model based on hydrodynamic simulations with Active Galactic Nucleus feedback \citep{battaglia:2010}, while \SPT\ adopts an Agora-based template \citep{agora}, both including only the late-time signal. Similarly, \PLK\ PR3 used a late-time model \citep{trac:2011}, whereas \hillipop\ PR4 combined late-time \citep{shaw:2012} and patchy \citep{battaglia:2013} spectra. More recently, \citet{gorce:2020} proposed models that describe the free-electron density contrast power spectrum in terms of reionisation history and morphology, allowing a natural combination of both contributions.

\begin{figure}[!ht]
	\centering
	\includegraphics[width=.85\columnwidth]{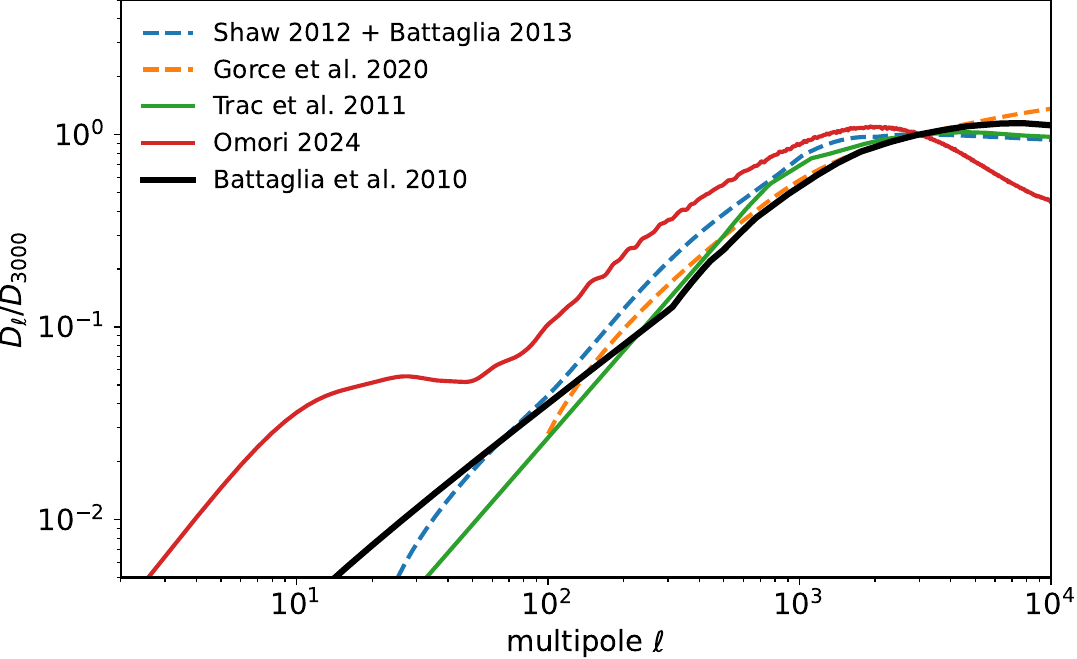}
	\caption{Templates of the kSZ power spectrum normalised at $\ell_0=3000$ (the black line corresponds to the baseline template). The late-time kSZ signal is modelled with solid lines, while dashed lines represent the combination of late-time and reionisation kSZ signals.}
	\label{fig:dl_ksz}
\end{figure}

We adopted the late-time kSZ of \citet{battaglia:2010} as a baseline and evaluated the impact of including a reionisation kSZ contribution and alternative template shapes (see Fig.~\ref{fig:dl_ksz}). The kSZ emission is parametrised by an amplitude $A^{\rm kSZ}$, defined at $\ell = 3000$, which rescales the template
\begin{equation}
\hat{C}_\ell^{\rm kSZ}(\nu\times\nu') = A^{\rm kSZ} \ \mathcal{C}_\ell^{\rm kSZ}.
\end{equation}

\subsection{Thermal \texorpdfstring{SZ$\times$CIB}{SZxCIB} correlation} 
Given our frequency range, we expect an anti-correlation between the tSZ signal from galaxy clusters and the CIB, since dark-matter overdensities host both hot ionised gas (producing negative tSZ) and DSFGs (positive CIB).

The $\ell$-dependence of this correlation has been modelled in different ways: as the geometric mean of the tSZ and CIB power spectra (as in the \SPT\ likelihood), through empirical descriptions calibrated to CIB simulations \citep{zahn:2012,shang:2012}, or using halo-model approaches \citep{addison:2012b,maniyar:2021}, the latter of which is adopted in the \PLK\ and \ACT\ likelihoods.

\begin{figure}[!ht]
	\centering
	\includegraphics[width=.85\columnwidth]{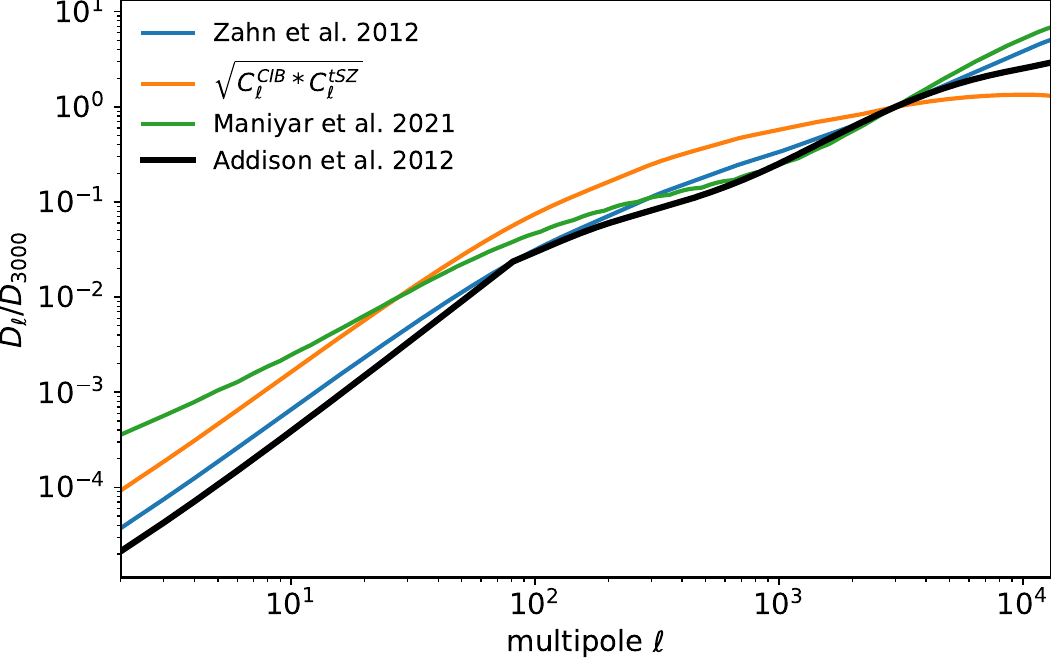}
	\caption{Templates of tSZ$\times$CIB power spectrum normalised at $\ell_0=3000$. The black line corresponds to the baseline template.}
	\label{fig:dl_szxcib}
\end{figure}

In our study, we adopted the \citet{addison:2012b} template as a baseline and tested alternative shapes (Fig.~\ref{fig:dl_szxcib}).
Given the amplitudes of the CIB and tSZ signals ($A^{\rm CIB}$, $A^{\rm tSZ}$), the correlation tSZ$\times$CIB is parametrised by 
\begin{align}
	\hat{C}_\ell^{\rm tSZ\times CIB}(\nu\times\nu') = {} & - \xi \sqrt{A^{\rm tSZ} A^{\rm CIB}} \nonumber\\
	& \times\ ( a_{\nu}^{\rm tSZ}a_{\nu'}^{\rm CIB} + a_{\nu}^{\rm CIB}a_{\nu'}^{\rm tSZ} )\ \mathcal{C}_\ell^{\rm tSZ\times CIB},
\end{align}
where $\xi$ denotes the correlation coefficient scaling the template $\mathcal{C}_\ell^{\rm tSZ\times CIB}$.

\subsection{Radio sources}
Unresolved radio sources are assumed to be Poisson distributed on the sky, resulting in a flat angular power spectrum. Their SED is usually modelled by a power law, $a_\nu^{\rm radio} = \nu^{-\beta_{\rm radio}}$, as in \citetalias{tristram:2024} and \citetalias{Louis:2025}. However, \citetalias{Camphuis:2025} uses a free amplitude parameter for each cross-frequency combination. The SED index $\beta_{\rm radio}$ is not well known and depends on the source population. 

In this study, we fitted $\beta_{\rm radio}$ and scaled the model with a free amplitude $A^{\rm radio}$, normalised at $\nu_0=150$\GHz, such that
\begin{equation}
    C_\ell^{\rm radio}(\nu\times\nu') = A^{\rm radio} \frac{a_{\nu}^{\rm radio}}{a_{\nu_0}^{\rm radio}} \frac{a_{\nu'}^{\rm radio}}{a_{\nu_0}^{\rm radio}} \, .
\end{equation}
We independently sampled the amplitude for each survey in $TT$ and $EE$. We assumed the contribution in $TE$ to be zero, as galaxy polarisation angles are uncorrelated. When we analysed individual datasets separately, we constrained $\beta_{\rm radio}$ with a Gaussian $\mathcal{N}(-0.8,0.1)$ \citep[see][]{tucci:2011}, otherwise we used a flat prior.

\section{Constraints on \texorpdfstring{$\mathsf{\Lambda}$CDM}{LCDM}}
\label{sec:lcdm}

In this section, we describe the constraints on the parameters of the \lcdm\ cosmological model with the likelihood described in Sect.~\ref{sec:lik}, using the three datasets: \PLK, \ACT, and \SPT.
We first present the results obtained using the baseline models, examining the consistency between the different datasets and between the spectra. We then discuss parameter correlations and explore the impact of foreground modelling on \lcdm\ parameters by varying the templates and model assumptions.

\begin{table*}[!ht]
    \caption{Parameter constraints for \lcdm.}
    \vspace{-0.25cm}
    \renewcommand{\arraystretch}{1.05}
    \center
    \begin{tabular}{l|ccc|c|c}
    \hline
    \hline
    Parameter     & \SPT\     &  \ACT\ & \PLK\ & \PLK+\ACT+\SPT  &  \PLK+\ACT+\SPT \\
    & (baseline) & (baseline) & (baseline) & (baseline) & (fg marg.) \\
    \hline
$\Omega_\mathrm{b}h^2$ & $0.02216 \pm 0.00021$ & $0.02258 \pm 0.00016$ & $0.02224 \pm 0.00014$ & $0.02228 \pm 0.00010$ & $0.02228 \pm 0.00010$\\
$\Omega_\mathrm{c}h^2$ & $0.1217 \pm 0.0038$ & $0.1234 \pm 0.0034$ & $0.1185 \pm 0.0012$ & $0.1196 \pm 0.0010$ & $0.1195 \pm 0.0010$\\
$100\theta_\mathrm{s}$ & $1.04167 \pm 0.00063$ & $1.04151 \pm 0.00037$ & $1.04184 \pm 0.00024$ & $1.04181 \pm 0.00018$ & $1.04182 \pm 0.00019$\\
$\mathrm{log}(10^{10}A_\mathrm{s})$ & $3.060 \pm 0.052$ & $3.068 \pm 0.047$ & $3.042 \pm 0.014$ & $3.053 \pm 0.013$ & $3.054 \pm 0.013$\\
$n_\mathrm{s}$ & $0.9475 \pm 0.0138$ & $0.9741 \pm 0.0089$ & $0.9701 \pm 0.0039$ & $0.9672 \pm 0.0034$ & $0.9669 \pm 0.0037$\\
$\tau$ & $0.0527 \pm 0.0297$ & $0.0651 \pm 0.0271$ & $0.0585 \pm 0.0063$ & $0.0611 \pm 0.0060$ & $0.0610 \pm 0.0060$\\
\hline
$H_0$ & $66.55 \pm  1.44$ & $66.29 \pm  1.24$ & $67.77 \pm  0.54$ & $67.42 \pm  0.39$ & $67.46 \pm  0.41$\\
$\Omega_\mathrm{m}$ & $0.3396 \pm 0.0229$ & $0.3469 \pm 0.0204$ & $0.3201 \pm 0.0075$ & $0.3258 \pm 0.0058$ & $0.3252 \pm 0.0061$\\
$\sigma_8$ & $0.8178 \pm 0.0172$ & $0.8330 \pm 0.0140$ & $0.8073 \pm 0.0068$ & $0.8142 \pm 0.0055$ & $0.8139 \pm 0.0056$\\
$S_8$ & $0.869 \pm 0.031$ & $0.895 \pm 0.023$ & $0.834 \pm 0.015$ & $0.848 \pm 0.010$ & $0.847 \pm 0.011$\\
    \end{tabular}
    \tablefoot{The last column shows the results of the chain marginalised over the foreground models. We report mean values and symmetrical 68\% confidence intervals.}
    \label{tab:lcdm}
\end{table*}

\subsection{\texorpdfstring{$\mathsf{\Lambda}$CDM}{LCDM}}
\label{sec:lcdm:dataset}

\begin{figure}[!ht]
	\centering
	\includegraphics[width=\columnwidth]{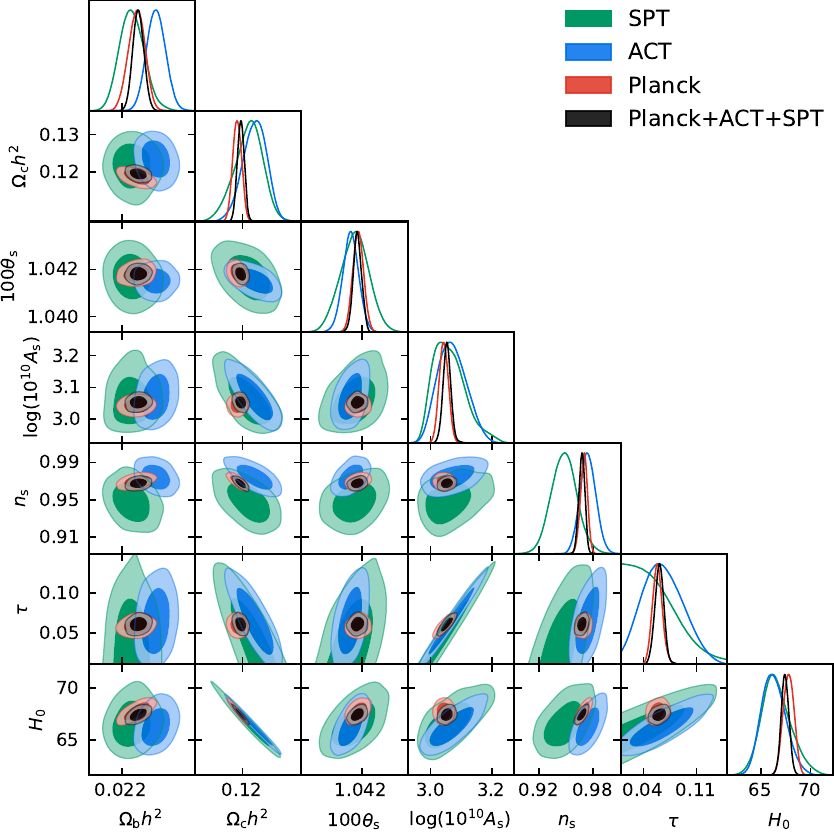}
	\caption{Posterior distributions for the \lcdm\ cosmological parameters derived from the \PLK, \ACT, and\SPT datasets and their combination, using the baseline foreground templates.}
	\label{fig:lcdm_3datasets}
\end{figure}

Figure~\ref{fig:lcdm_3datasets} presents a comparison of the posterior distributions for the \lcdm\ parameters derived from \PLK, \ACT, \SPT,  and their combination, using the coherent likelihood described in Sect.~\ref{sec:lik} and the baseline foreground templates (Sect.~\ref{sec:model}). Each dataset incorporates measurements from the $TT$, $TE$, and $EE$ power spectra. Table~\ref{tab:lcdm} summarises constraints on \lcdm\ cosmological parameters.

Overall, the three datasets show very good agreement on the \lcdm\ parameters.
We observe a slight deviation for $\Omega_{\rm b} h^2$ between \ACT\ and the other datasets, although the difference remains only at the 1.8$\sigma$ level.
We also note a slightly lower value of the scalar index $n_{\rm s}$ for \SPT, although it is also below $1.5\sigma$.
Due to the absence of low-multipole data in \ACT\ and \SPT, the degeneracy between $\tau$ and $A_{\rm s}$ remains unresolved, resulting in large uncertainties for these two parameters.
Nevertheless, the overall level of consistency is sufficient to justify combining the datasets.

The constraints on \lcdm\ parameters are dominated by \PLK, although \ACT\ approaches the sensitivity of \PLK for $\Omega_{\rm b}h^2$.
Relative to \PLK\ alone, the joint analysis of the three CMB datasets yields substantial reductions in the uncertainties in $\Omega_{\rm b}h^2$, $\Omega_{\rm c}h^2$, and $\theta_{\rm s}$ by 50\%, 17\%, and 22\%, respectively (see Table~\ref{tab:lcdm}). In contrast, the improvements for $\tau$, $A_{\rm s}$, and $n_{\rm s}$ are marginal.
When all datasets are combined, the uncertainty on the amplitude of scalar perturbations, $\log(10^{10}A_{\rm s})$, is slightly larger than that reported in \citetalias{Camphuis:2025} ($\sigma_{\log(10^{10}A_{\rm s})} = 0.0099$), but is compatible with the value in \citetalias{Louis:2025} ($\sigma_{\log(10^{10}A_{\rm s})} = 0.013$). This increase arises because we relaxed some constraints on the absolute calibration and allowed the polarisation efficiencies to vary freely.

Correlations between foreground and cosmological parameters are generally weak, supporting the robustness of the cosmological results with respect to the foreground modelling (see Fig.~\ref{fig:corr}). Although small, these correlations are not strictly zero and motivate a closer investigation of the impact of foreground modelling on the \lcdm\ parameters (Sect.~\ref{sec:lcdm:fgmarg}).
In contrast, we observe strong correlations among the extragalactic foreground parameters, whose impact is further explored in Sect.~\ref{sec:foregrounds}.

We do not expect the results presented here to exactly match those obtained by the respective teams, mainly because of differences in the foreground modelling and in the adopted priors. For cosmology, we did not apply any prior on the reionisation optical depth $\tau$. For instrumental systematics, we adopted the priors specific to each experiment. For the foreground components, we did not use informative priors, except for the dust amplitudes in the \ACT\ and \SPT\ datasets. The only additional exceptions are $\beta_{\rm CIB}$ and $\beta_{\rm radio}$, for which we applied Gaussian priors when analysing the datasets individually.
We verified that, using the same assumptions, we recovered the same results published in \citetalias{tristram:2024}, \citetalias{Louis:2025}, and \citetalias{Camphuis:2025} for \PLK, \ACT, and \SPT, respectively.

\subsection{Consistency between temperature and polarisation}
\label{sec:lcdm:modes}

\begin{figure}[!ht]
	\centering
	\includegraphics[width=\columnwidth,height=210px]{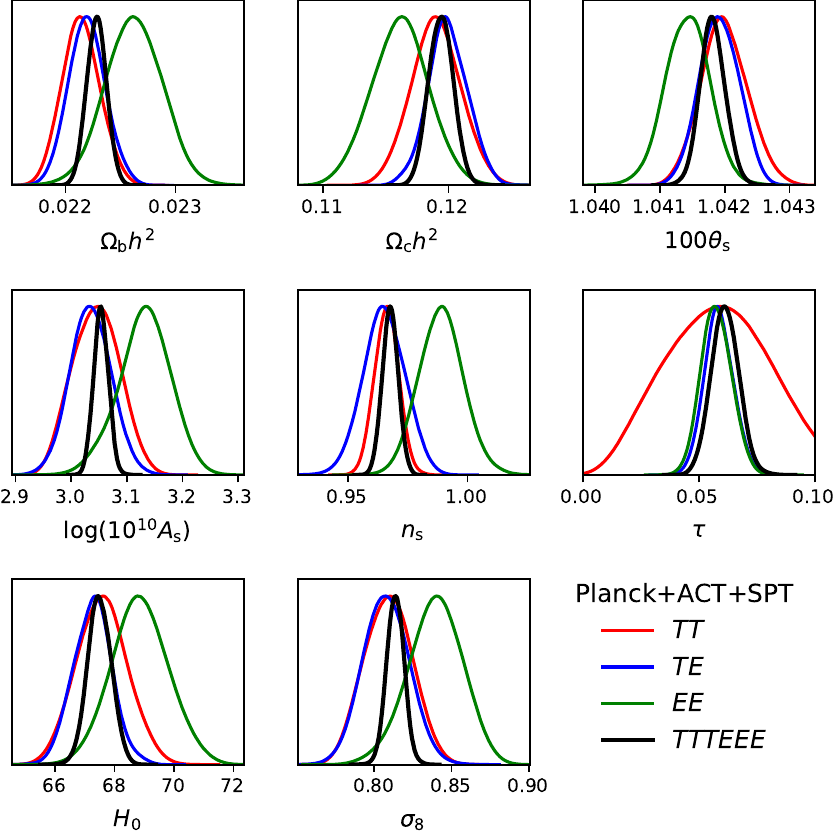}
	\caption{Posterior distributions for the \lcdm\ cosmological parameters derived from the combination of \PLK, \ACT, and \SPT, using $TT$, $TE$, $EE$ power spectra and their combination.}
	\label{fig:lcdm_TT_TE_EE}
\end{figure}

We checked the agreement of \lcdm\ parameter constraints from the temperature and polarisation data. This allowed us to assess the ability of \lcdm\ to jointly describe temperature and polarisation data. 
For $EE$ and $TE$, we included the \lollipop\ low-$\ell$ $EE$ likelihood from \PLK. For $TT$, we included the \commander\ low-$\ell$ $TT$ likelihood.
We find very good consistency between the constraints from $TT$ and $TE$, while those from $EE$ are broader and show a mild shift in the amplitude $A_{\rm s}$ and the index $n_{\rm s}$ of the scalar fluctuations towards higher values (Fig.~\ref{fig:lcdm_TT_TE_EE}). 
In general, the $TT$ and $TE$ spectra provide comparable constraining power, but the relative weight depends on the parameter: for example, the combined result for $\Omega_{\rm c}h^2$ is primarily driven by $TE$, whereas the constraint on $n_{\rm s}$ is dominated by $TT$.

\subsection{Impact of foregrounds models}
\label{sec:lcdm:fgmarg}

\begin{figure}[htbp!]
	\centering
	\includegraphics[width=\columnwidth,height=140px]{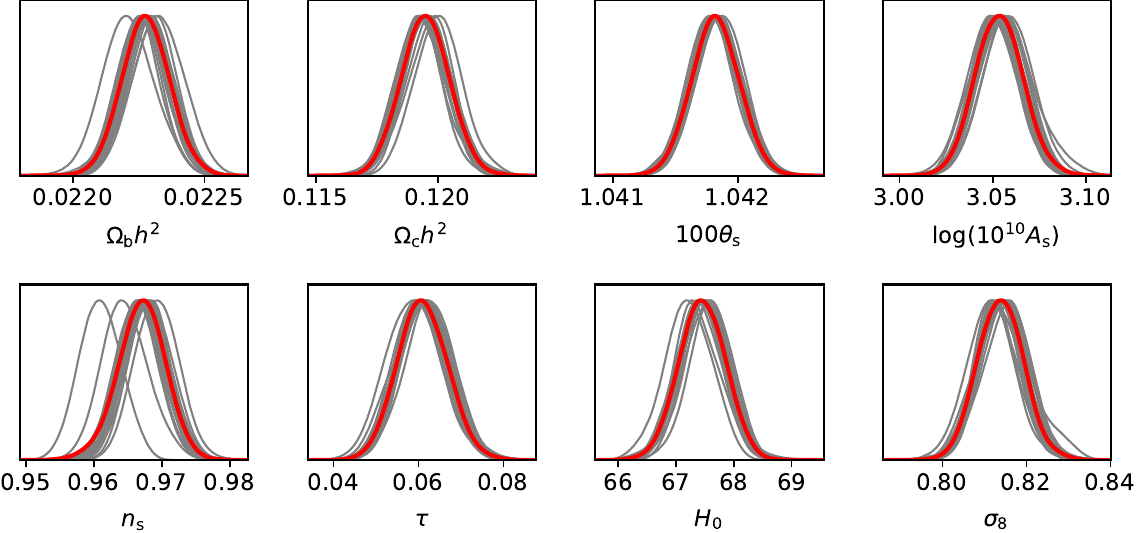}
	\caption{Posterior distributions for the \lcdm\ parameters obtained when varying templates for the foreground models as described in Sect.~\ref{sec:foregrounds} (in grey), compared with the posteriors after marginalisation over the foreground models (in red).}
	\label{fig:lcdm-templates}
\end{figure}

Figure~\ref{fig:lcdm-templates} illustrates how the posterior distributions of the \lcdm\ parameters change when different foreground templates are considered.
We ran several chains corresponding to different combinations of foreground spectra to assess how parameter estimates vary with foreground modelling details. In practice, we varied only one foreground model at a time and then concatenated the resulting MCMC chains into a pseudo-chain from which we drew the marginal posteriors. 
This procedure approximates Bayesian model averaging under the assumption that the models considered are equally probable and yield nearly identical Bayesian evidence. Indeed, the maximum variation of $\chi^2$ across the models is less than $26.4$ for a data vector of size 7403 (see Tab.~\ref{tab:chi2}). In this regime, the pseudo-chain posteriors provide a good proxy for full marginalisation over foreground models.

Overall, template choice has only a minor impact, confirming that the cosmological results are robust against reasonable variations in foreground modelling. As reported in Table~\ref{tab:lcdm}, marginalisation over the foreground templates increases parameter uncertainties by no more than 5\%, with the exception of $n_{\rm s}$, for which the uncertainty increases by 9\%.
The mean value of $n_{\rm s}$ is slightly affected by assumptions on the dust and point-source models, reflecting the mild but non-negligible correlation observed between $n_{\rm s}$ and the clustered CIB amplitude, $A_{\rm CIB}$, in the parameter correlation matrix. In particular, the main outlier in the $n_{\rm s}$ posterior distributions corresponds to the case where the greybody assumption for the dust SED is relaxed and independent dust amplitudes are fitted for each frequency and mode, which drives $A_{\rm CIB}$ higher and $n_{\rm s}$ lower.

\section{Discussions on the foregrounds}
\label{sec:foregrounds}

\begin{figure}[!ht]
	\centering
	\includegraphics[width=\columnwidth]{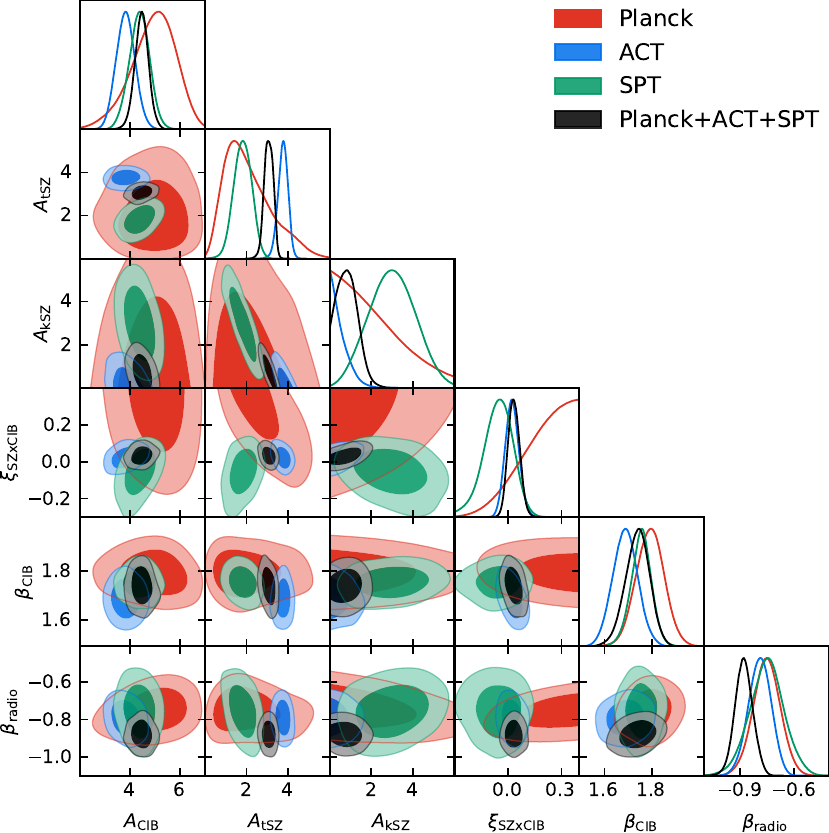}
	\caption{Posterior distributions for the extragalactic parameters from \PLK, \ACT, and \SPT.}
	\label{fig:par_extragal}
\end{figure}

An important aspect of this work is the adoption of a coherent foreground model across all likelihoods, ensuring that the different datasets can be compared and combined on equal footing. Figure~\ref{fig:par_extragal} presents the posterior distributions for the amplitudes of the CIB, tSZ, and kSZ components, together with the correlation coefficient $\xi_\mathrm{tSZ\times CIB}$ and the SED indices for the CIB and radio sources, shown for each experiment individually and for the combined analysis. 
Compared to the tight constraints on the \lcdm\ parameters, \PLK\ alone provides weaker constraints on the foreground sector, while \ACT\ and \SPT\ add significant constraining power. 

In this section, we further exploit the flexibility of our likelihood framework to explore the impact of foreground modelling, both to assess the robustness of the cosmological constraints and to investigate how the recovered foreground parameters depend on the choice of the model.
For each extragalactic foreground amplitude, we computed the average of the mean posterior across all foreground model combinations and report two types of uncertainties: the average of the recovered uncertainties (labelled \textit{stat}) and the dispersion of the mean arising from variations in the full set of foreground models (labelled `fg').

\subsection{Galactic foregrounds}
\label{sec:foregrounds:dust}

We explored different assumptions for modelling the dust residual component. First, we removed the prior on the dust SED index $\beta_{\rm d}$. Then, we opened the priors on the dust amplitude $A_{\rm dust}$ associated with \ACT\ and \SPT. Finally, we considered a parametrisation without an SED and fitted one amplitude for each cross-spectrum.

\begin{figure}[!ht]
	\centering
	\includegraphics[width=.9\columnwidth]{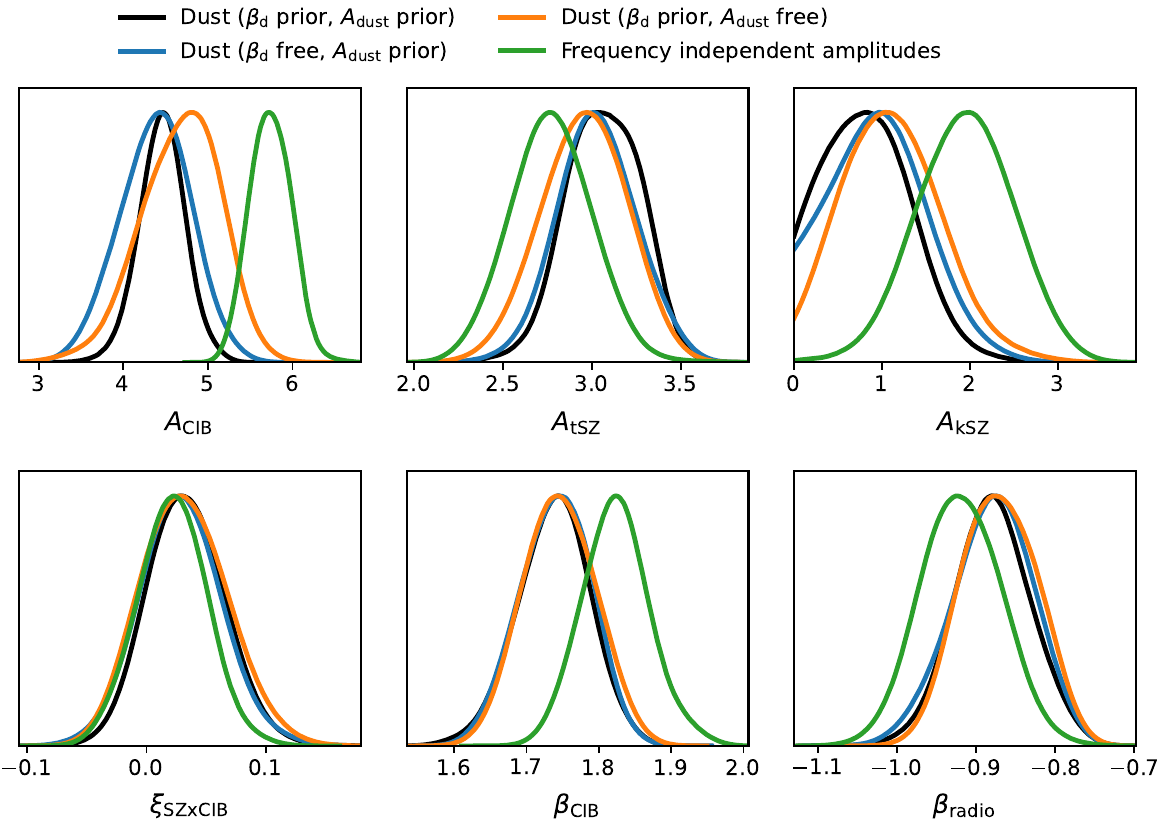}
	\caption{Posterior distributions for extragalactic foregrounds under various hypotheses on the dust model for the combined \PLK, \ACT, and \SPT\ dataset.}
	\label{fig:fg_dust}
\end{figure}

The constraint on the dust SED spectral index is driven by the priors $\beta^{TT}_{\rm d} = \mathcal{N}(1.51,0.01)$ and $\beta^{TE/EE}_{\rm d} = \mathcal{N}(1.59,0.02)$, as the data lack the sensitivity to measure this parameter directly. 
We verified this by removing the priors: we find the same results for all extragalactic foregrounds, with \text{$\beta_{\rm d}^{TT} = 1.44 \pm 0.17$} and essentially flat posteriors for $\beta^{TE/EE}_{\rm d}$, confirming that the datasets themselves provide no meaningful constraint.

The recovered constraint on the power-law index of the dust template in temperature is 
$$\alpha_{\rm d}^{TT} = -2.620 \pm 0.054\text{(stat)} \pm 0.037\text{(fg)},$$ 
in agreement with the \PLK\ measurement, and is primarily driven by the \PLK\ data. In polarisation, the sensitivity is lower than in temperature, so we adopted the \PLK\ fit and fixed $\alpha_{\rm d} = -2.4$ for $TE$ and $EE$. Allowing these indices to vary yields consistent results, with $\alpha_{\rm d}^{TE} = -2.50 \pm 0.09$ and $\alpha_{\rm d}^{EE} = -2.40 \pm 0.06$.

Removing the priors on the dust amplitude (orange curve in Fig.~\ref{fig:fg_dust}) yields very similar posteriors for all the extragalactic foregrounds. However, it significantly slows down the MCMC convergence because the two ground-based observatories provide little direct information on the dust parameters.

Finally, avoiding a specific dust model requires fitting the amplitudes of the $T$ and $E$ signals independently at each frequency. This introduces additional degeneracies with other foregrounds and shifts the extragalactic parameters, noticeably increasing the CIB amplitude (green curve in Fig.~\ref{fig:fg_dust}).

Overall, the impact of dust modelling is most significant for the reconstruction of the CIB signal. This is because the spectral energy distributions of Galactic dust and CIB are very similar, leading to correlations between the two components. As a result, uncertainties in the dust model propagate directly into the inferred CIB amplitude.

\subsection{Cosmic infrared background}
\label{sec:foregrounds:cib}

\begin{figure}[!ht]
	\centering
	\includegraphics[width=.9\columnwidth]{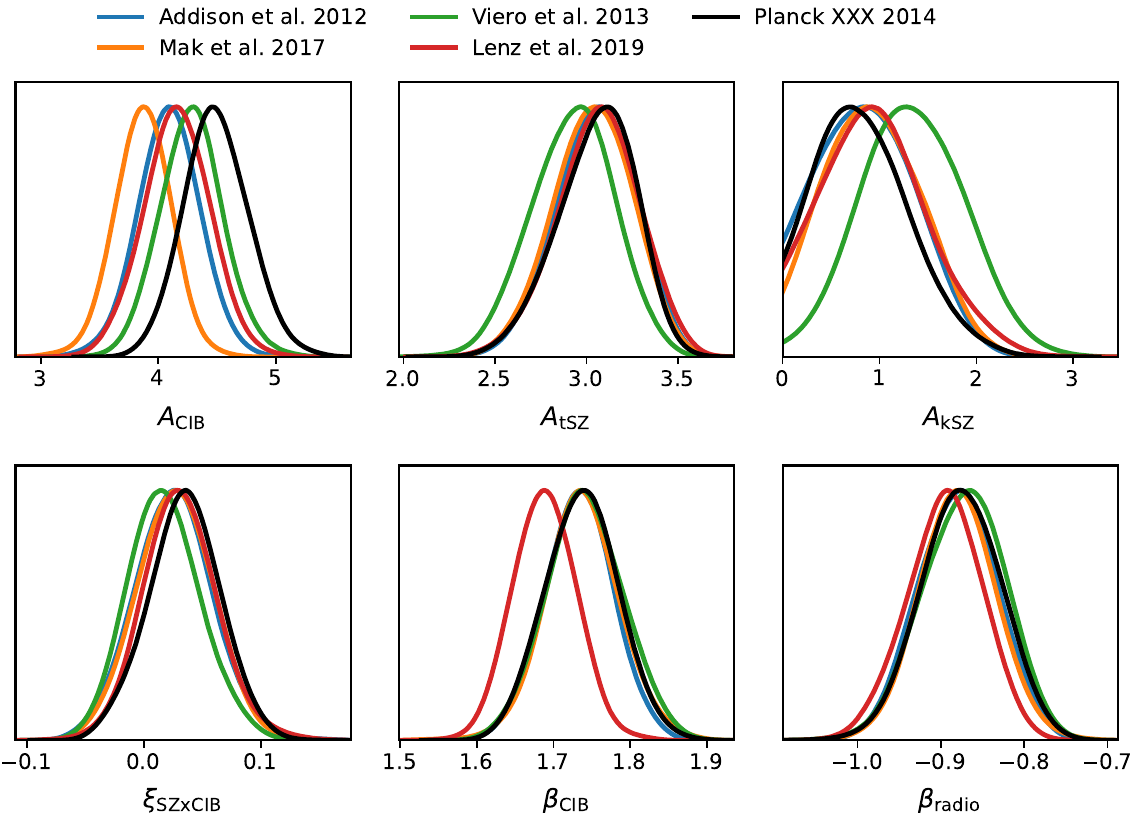}
	\caption{Posterior distributions for extragalactic foreground using the CIB templates shown in Fig.~\ref{fig:dl_cib} for the combined \PLK, \ACT, and \SPT\ dataset.}
	\label{fig:fg_cib}
\end{figure}

We find a clear detection of the CIB, with an amplitude at $\ell=3000$ well measured by the three datasets. However, the amplitude of the CIB signal depends significantly on the model considered, as shown in Fig.~\ref{fig:fg_cib}.

The recovered value for the amplitude of the CIB is 
$$A_{\rm CIB} = 4.32 \pm 0.33 \text{(stat)} \pm 0.60 \text{(fg)}\,\mu\text{K}^2,$$ 
where the dispersion of the mean due to foreground modelling is about twice the statistical uncertainty.
This detection is very close to the result of the combined datasets in \citetalias{Beringue:2025}, $3.93 \pm 0.26\ \mu\mathrm{K}^2$, is compatible with \ACT, $3.69 \pm 0.47\ \mu\mathrm{K}^2$ \citepalias{Louis:2025}, but significantly exceeds \SPT, which finds $1.88 \pm 0.80\ \mu\mathrm{K}^2$ for the $150\times150$ cross-spectrum \citepalias{Camphuis:2025}.

For the index of the CIB spectral energy density, we find 
$$\beta_{\rm CIB} = 1.79 \pm 0.07\text{(stat)} \pm 0.16\text{(fg)},$$ 
in good agreement with independent measurements from \PLK\ CIB reconstructed maps, $\beta_{\rm CIB}=1.75\pm0.06$ \citep{planck2013-pip56}. Once again, uncertainty associated with the choice of foreground templates dominates over the statistical error.

\subsection{Thermal Sunyaev--Zeldovich}
\label{sec:foregrounds:tsz}

The tSZ signal is detected with high significance. 
However, we observe a substantial dispersion in the recovered posterior distributions across datasets (see Fig.~\ref{fig:par_extragal}), with amplitudes ranging from $\sim$2 to 4\muK$^2$.

\begin{figure}[!ht]
	\centering
	\includegraphics[width=0.9\columnwidth]{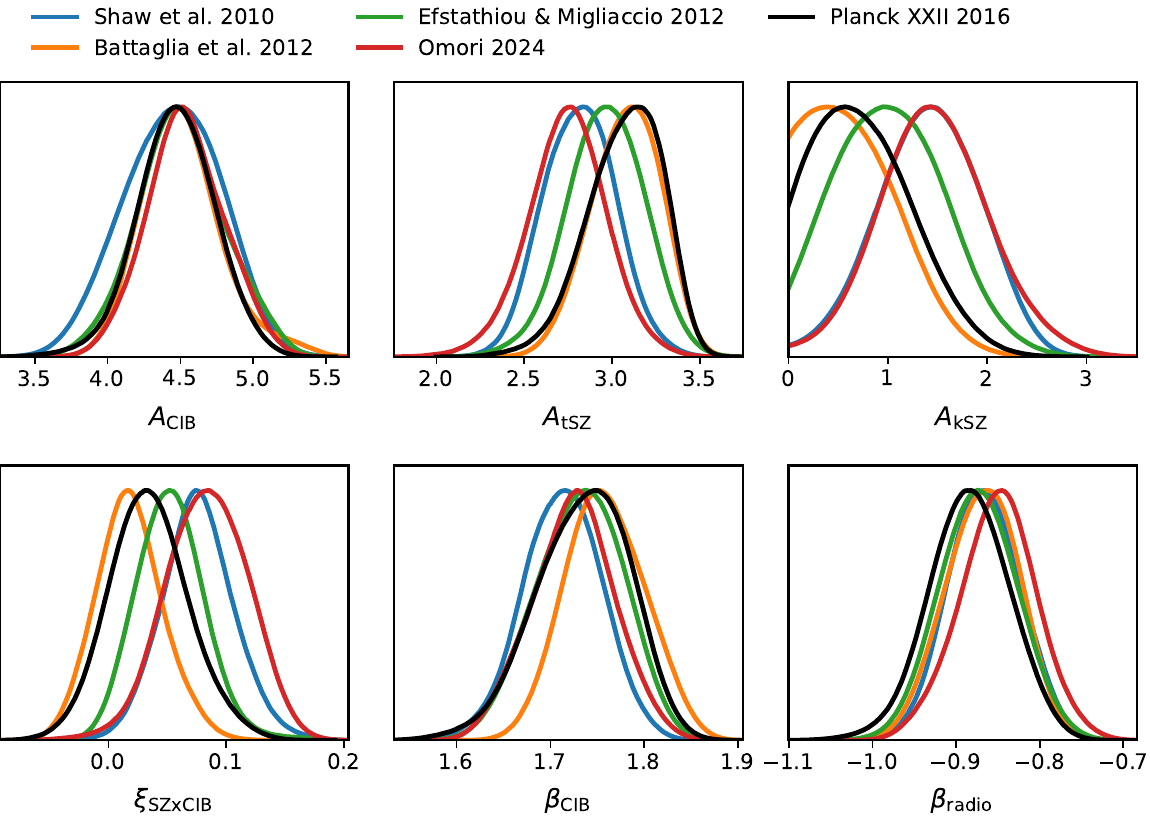}
	\caption{Posterior distributions for extragalactic foregrounds with the tSZ templates shown in Fig.~\ref{fig:dl_tsz} for the combined \PLK, \ACT, and \SPT dataset.}
	\label{fig:fg_tsz}
\end{figure}

Varying the tSZ template changes the amplitude from 2.7 to $3.3\muK^2$ (see Fig.~\ref{fig:fg_tsz}). We highlight that both the amplitude of the kSZ component and the tSZ$\times$CIB correlation are also affected by the choice of the tSZ template.
Combining the three datasets, we find
$$A_{\rm tSZ} = 2.98 \pm 0.22 \text{(stat)} \pm 0.10 \text{(fg)}\,\mu\text{K}^2,$$ 
where the dispersion of the mean due to foreground modelling corresponds to approximately 50\% of the statistical uncertainty. 

This result is consistent with the value reported by \citetalias{Beringue:2025} ($3.46^{+0.29}_{-0.25}\ \mu\mathrm{K}^2$), or ACT ($3.35 \pm 0.35\ \mu\mathrm{K}^2$, \citetalias{Louis:2025}), but less compatible with that of SPT, \text{$0.93 \pm 0.5\ \mu\mathrm{K}^2$} \citepalias{Camphuis:2025}. The lower amplitude recovered in the \SPT\ dataset may be due to the masking of the brightest clusters for the estimation of the angular power spectra. This suggests that both the amplitude and shape of the tSZ contribution may need to be adjusted depending on the cluster masking strategy.

Taking into account an additional scaling parameter, with all datasets together, we find no evidence of a deviation from the baseline shape, obtaining $\alpha_{\rm tSZ} = -0.11 \pm 0.13$.

\subsection{Kinetic Sunyaev--Zeldovich}
\label{sec:foregrounds:ksz}

\begin{figure}[!ht]
	\centering
    \includegraphics[width=0.9\columnwidth]{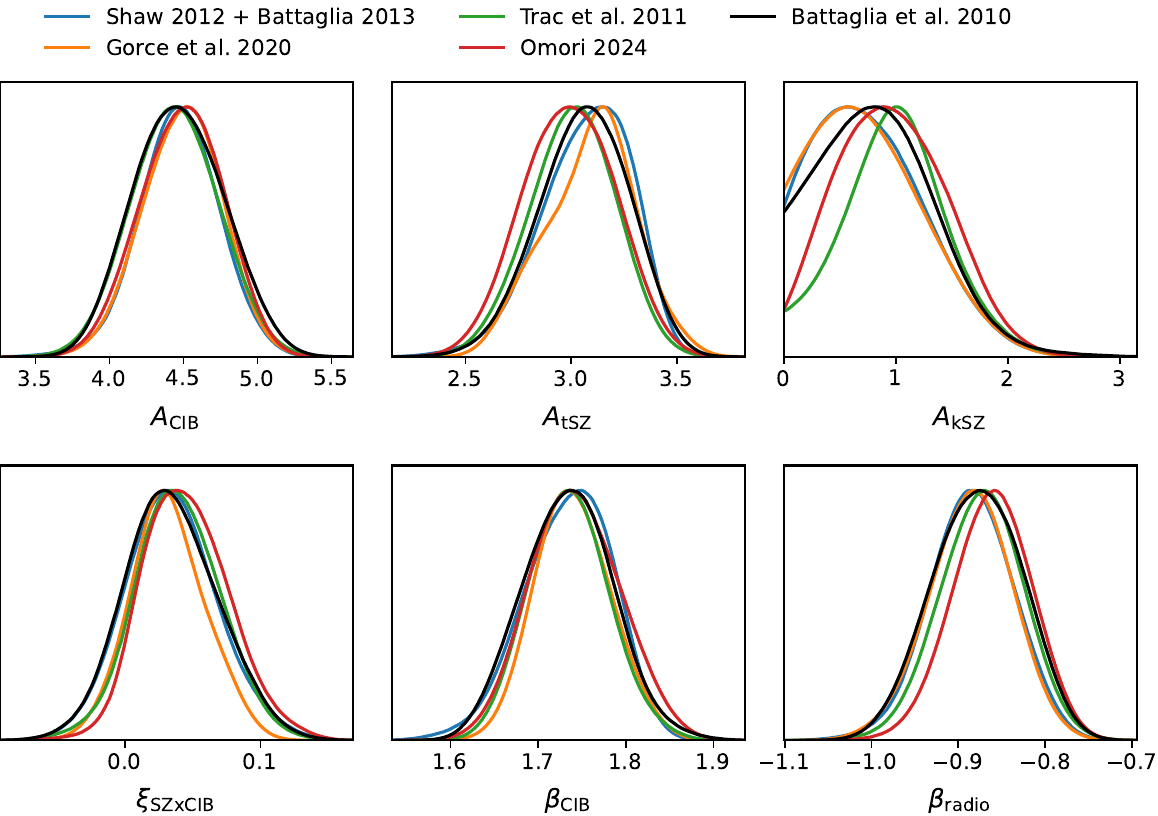}
	\caption{Posterior distributions for extragalactic foregrounds with the kSZ templates shown in Fig.~\ref{fig:dl_ksz} for the combined \PLK, \ACT, and \SPT\ dataset.}
	\label{fig:fg_ksz}
\end{figure}

The kSZ is not detected by \PLK\ and \ACT, but we find some signal in the \SPT\ data around $3\muK^2$ (Fig.~\ref{fig:par_extragal}). When we combine the three datasets, our result remains compatible with zero and stable across different kSZ power-spectrum templates (Fig.~\ref{fig:fg_ksz}). However, the recovered amplitude of the kinetic SZ signal varies significantly with the assumptions adopted for other foreground components, particularly the tSZ (Fig.~\ref{fig:fg_tsz}), the tSZxCIB correlation (Fig.~\ref{fig:fg_szxcib}), and the point sources (Fig.~\ref{fig:fg_ps}).

Across all templates, we find upper limits at the 95\% confidence level ranging from $A_\mathrm{kSZ} < 1.43$ to $5.36\muK^2$.
This can be compared with the results from\ACT, which report an upper limit at 95\% CL of $A_\mathrm{kSZ} < 3.7\muK^2$ \citepalias{Louis:2025} and from \SPT, for which \citetalias{Camphuis:2025} find $A_\mathrm{kSZ} < 2.90\muK^2$).
Averaging over all foreground models, we find a kSZ amplitude of 
$$A_{\rm kSZ} = 1.20 \pm 0.52 \text{(stat)} \pm 0.64 \text{(fg)}\,\mu\text{K}^2,$$
where the uncertainty arising from foreground templates exceeds the statistical one. It is compatible with the value from \citetalias{Beringue:2025}, $1.34^{+0.57}_{-0.71}\,\mu\text{K}^2$. When marginalising over the foreground modelling, the result is consistent with zero at the $1.4\sigma$ level, with an upper-limit of $A_\mathrm{kSZ} < 2.79\muK^2$ (95\% CL).

\subsection{tSZxCIB}
\label{sec:foregrounds:szxcib}

\begin{figure}[!ht]
	\centering
	\includegraphics[width=0.9\columnwidth]{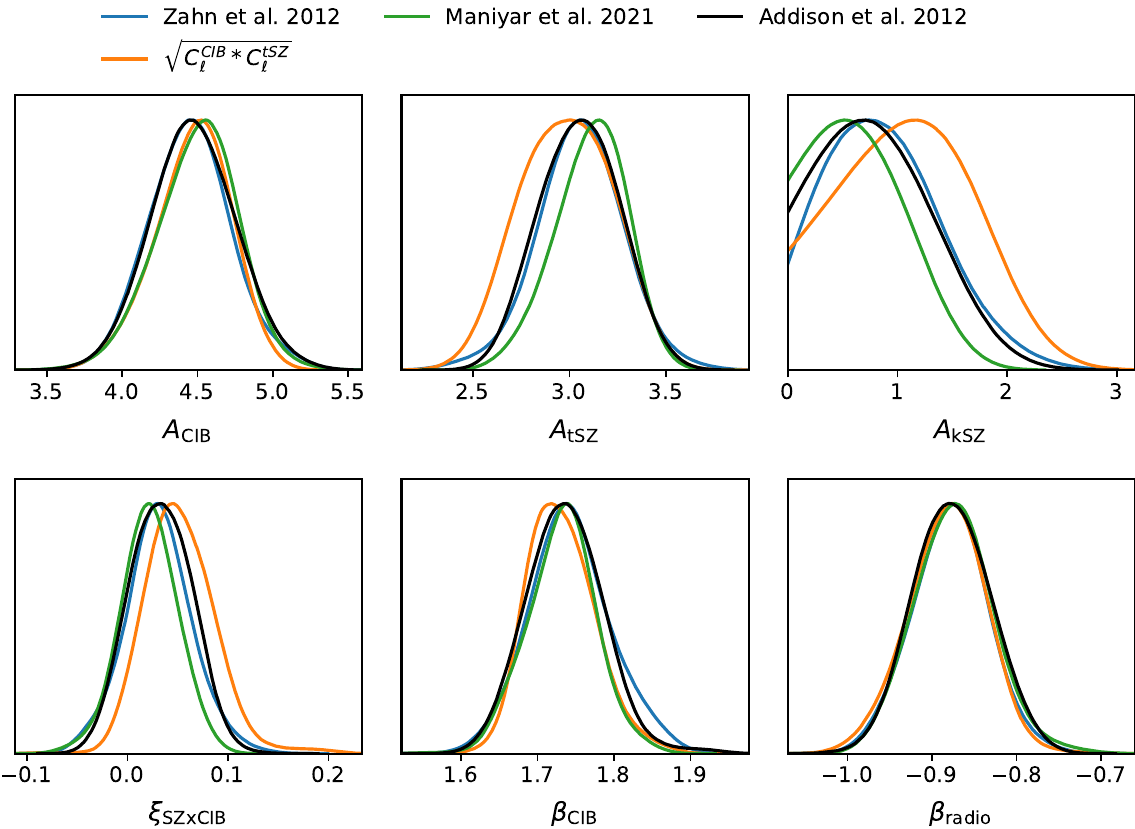}
	\caption{Posterior distributions for extragalactic foregrounds with the tSZ$\times$CIB templates shown in Fig.~\ref{fig:dl_szxcib} for the combined \PLK, \ACT, and \SPT\ dataset.}
	\label{fig:fg_szxcib}
\end{figure}

We find that the cross-correlation between tSZ and CIB is consistent with zero for all datasets (Fig.~\ref{fig:par_extragal}). 
The correlation coefficient $\xi$ is affected by the shape of the tSZ signal (Fig.~\ref{fig:fg_tsz}) and the point-source model (Fig.~\ref{fig:fg_ps}), but we find only a weak dependence on the shape of the tSZ$\times$CIB model (Fig.~\ref{fig:fg_szxcib}).

Across all foreground models, the amplitude of the tSZ$\times$CIB correlation coefficient is 
$$\xi_{\rm tSZ \times CIB} = 0.044 \pm 0.036\text{(stat)} \pm 0.045\text{(fg)}.$$
The constraint is compatible with that of \ACT, $\xi = 0.088^{+0.036}_{-0.075}$ \citepalias{Louis:2025}, and is consistent with the \SPT\ analysis that neglects this signal.

\subsection{Point sources}
\label{sec:foregrounds:ps}

We independently fitted the amplitudes of radio sources for each dataset to account for the different flux cuts of each survey, while we assumed a common spectral index for the SED. We recover 
$$\beta_{\rm radio} = -0.877 \pm 0.048\text{(stat)} \pm 0.013\text{(fg)},$$ 
consistent with expectations from the source studies \citep{tucci:2011,Lagache:2020}.

When fitted separately from the CIB, we find the index of the Poisson term of the CIB, corresponding to DSFGs, to be $\beta_{\rm DSFG} = 1.68 \pm 0.06$, with an associated clustered CIB index of $\beta_{\rm CIB} = 2.06 \pm 0.16$. For comparison, when we used the same index for both Poisson and clustered terms, we find $\beta_{\rm CIB} = 1.79 \pm 0.19$ after marginalisation (Sect.~\ref{sec:foregrounds:cib}).

\begin{figure}[!ht]
	\centering
	\includegraphics[width=0.9\columnwidth]{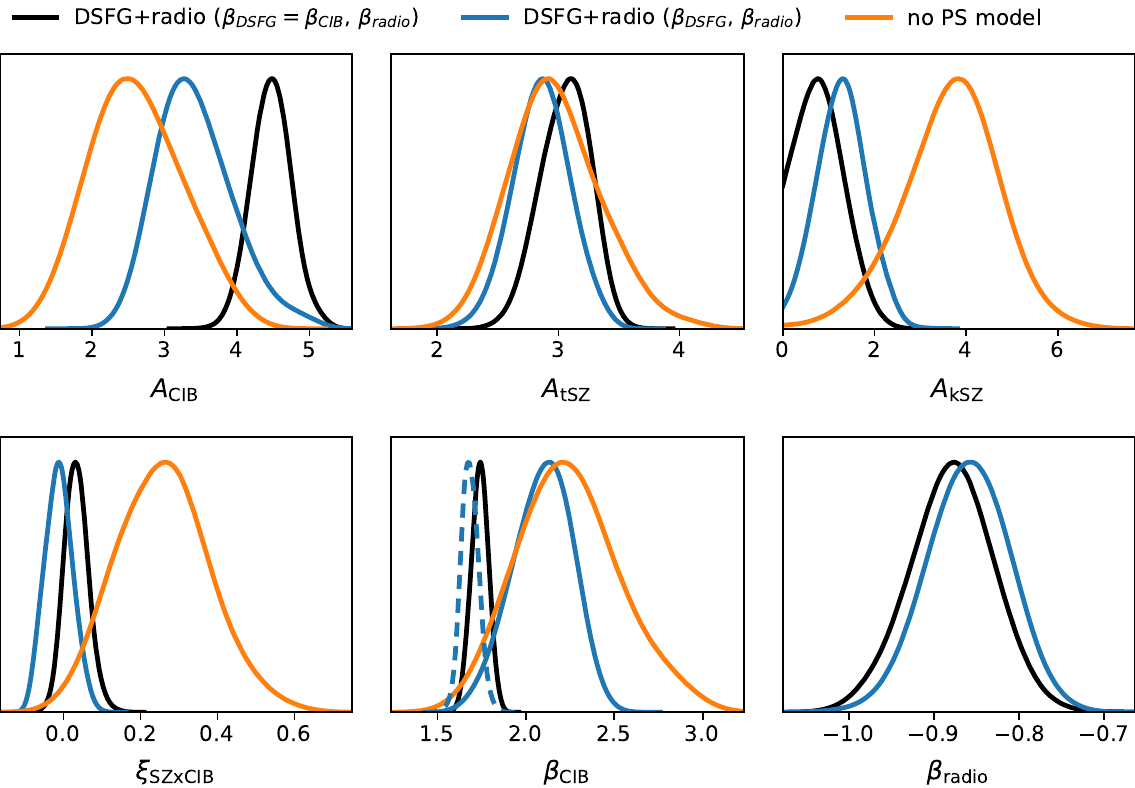}
	\caption{Posterior distributions for extragalactic foregrounds using various point sources models. The dashed line indicates the DSFG index ($\beta_{\rm DSFG}$).}
	\label{fig:fg_ps}
\end{figure}

We also tested how the recovered foreground parameters change when point sources are not jointly modelled but are instead fitted independently for each cross-spectrum of each dataset (see the orange curve in Fig.~\ref{fig:fg_ps}). Because the number of free parameters is much larger (18 instead of seven in the point-source sector), the resulting uncertainties are correspondingly larger.
Nevertheless, the impact of point-source modelling is significant: it can suppress the CIB amplitude by more than a factor of two while simultaneously enhancing the kSZ signal and the tSZ$\times$CIB correlation.

\section{Extensions to \texorpdfstring{$\mathsf{\Lambda}$CDM}{LCDM}}
\label{sec:extensions}

We explored several extensions to the baseline \lcdm\ model -- namely the phenomenological lensing amplitude \Alens, the effective number of relativistic species \Neff, the total neutrino mass \Mnu, and the spatial curvature \Ok -- which have been at the centre of key CMB tensions in recent years. Testing these parameters within our joint framework is particularly informative, as combining \PLK\ with high-resolution, small-scale measurements from \ACT\ and \SPT\ helps break degeneracies (see Annex~\ref{ann:ext}) and assess whether these anomalies persist when the datasets are modelled consistently.

\begin{figure*}[!ht]
    \centering
	\includegraphics[width=.3\textwidth,height=120pt]{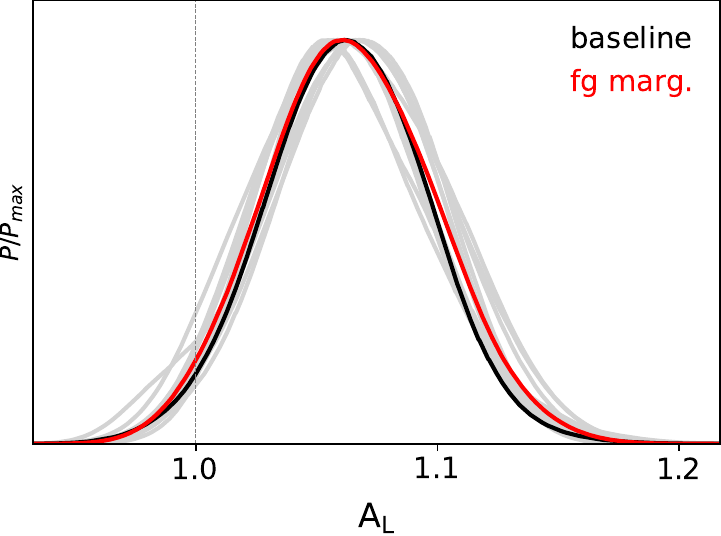}
    \hspace{10pt}
	\includegraphics[width=.3\textwidth,height=120pt]{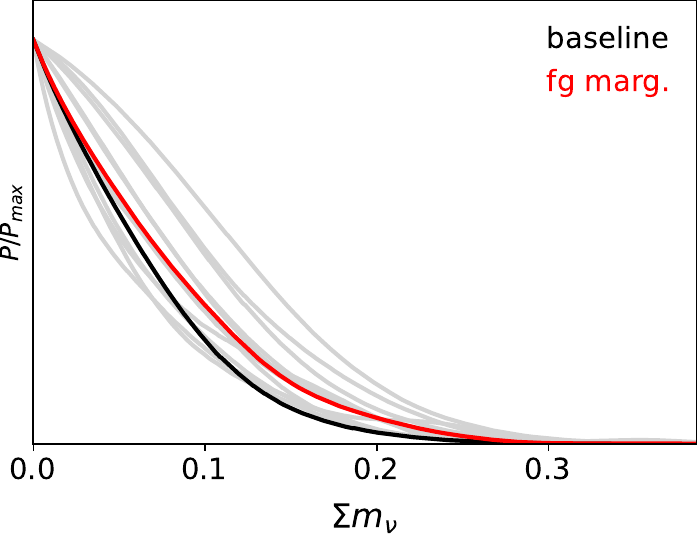}
    \hspace{10pt}
	\includegraphics[width=.3\textwidth,height=120pt]{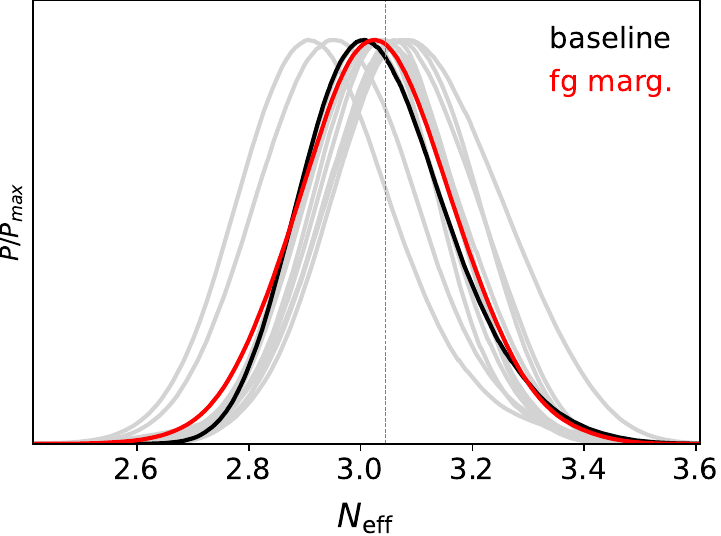}
	\caption{Posterior distributions for \Alens\ (\textit{left}), \Mnu\ (\textit{center}), and \Neff\ (\textit{right}) combining \PLK, \ACT, and \SPT, with varying foreground templates as described in Sect.~\ref{sec:foregrounds} (\textit{in grey}), compared with the posteriors after marginalisation over the foreground models (\textit{in red}).}
	\label{fig:ext_marg}
\end{figure*}

\subsection{Amplitude of the gravitational lensing, \texorpdfstring{\Alens}{Alens}}

The phenomenological parameter \Alens\ rescales the amplitude of the gravitational lensing potential in the CMB power spectra, effectively modulating the amount of peak smoothing induced by lensing. Previous analyses of \PLK\ data report values of $\Alens>1$, suggesting an apparent excess of lensing compared to \lcdm\ predictions and motivating further investigation with independent datasets.

Combining the three CMB datasets and averaging over the foreground models, we obtain 
$$\Alens = 1.063 \pm 0.033\,\text{(stat)} \pm 0.003\,\text{(fg)} \quad \text{(\PLK+ACT+SPT)}.$$
This result is slightly lower and more precise than the values reported in \citetalias{Camphuis:2025} for the same dataset combination (\text{$\Alens = 1.083 \pm 0.037$}) or in the \PACT\ combination of \citetalias{Louis:2025} (\text{$\Alens = 1.081 \pm 0.043$}).

Overall, we find a limited impact of the foreground modelling on the estimation of $\Alens$, and no evidence of anomalous lensing relative to \lcdm\ expectations.

\subsection{Sum of the neutrino masses, \texorpdfstring{\Mnu}{Mnu}}
Figure~\ref{fig:ext_marg} shows the posterior distributions for the sum of the neutrino masses, \Mnu. 
Averaging over the different foreground models, we obtain 
$$\Mnu < 0.181\,\text{eV} \quad \text{(95\% CL, \PLK+\ACT+\SPT)},$$
with a standard deviation of $0.019\,\text{eV}$ across foreground models. This dispersion represents 35\% of the 1-$\sigma$ statistical sensitivity ($0.055\,\text{eV}$), indicating that the combined constraint is sensitive to the details of the foreground modelling.

Compared to the \PLK-only limit (see Annexe~\ref{ann:ext}), the combined analysis improves the upper bound by more than a factor of two. The improvement is primarily driven by the small-scale information provided by \ACT\ and \SPT, which are sensitive to the modification of smoothing due to gravitational lensing in the $TT$ and $EE$ power spectra.

\subsection{Effective number of relativistic species, \texorpdfstring{\Neff}{Neff}}
The effective number of relativistic species, \Neff, parametrises the total radiation energy density in the early Universe and is sensitive to any additional light relics beyond the three active neutrino species predicted by the standard model.

Figure~\ref{fig:ext_marg} shows the posterior distributions for the combination  of \PLK, \ACT, and \SPT\ for the various foreground templates considered in this paper. Averaging over foreground models, 
$$\Neff = 3.031 \pm 0.130\,\text{(stat)} \pm 0.045\,\text{(fg)} \quad \text{(\PLK+\ACT+\SPT)}.$$
This result is in excellent agreement with the standard model expectation of $\Neff = 3.044$ for three neutrino species undergoing non-instantaneous decoupling from the primordial plasma \citep{froustey:2020,akita:2020}.

This combined result is more precise than any individual dataset and highlights the importance of both the improved \PLK\ PR4 data release and the use of a coherent joint likelihood. In contrast, earlier analyses based on PR3 data and obtained by only multiplying likelihoods reported lower values, such as $\Neff = 2.81 \pm 0.12$ in \citetalias{Camphuis:2025} and $\Neff = 2.73 \pm 0.14$ in \citet{Calabrese:2025}. With PR4 and the present methodology, we no longer observe these apparent downward shifts: our result is fully consistent with the standard model expectation and robust against the foreground modelling.

\subsection{Spatial curvature, \texorpdfstring{\Ok}{Ok}}
The curvature parameter, \Ok, measures the deviation of the Universe from spatial flatness. Analyses based on \PLK\ PR3 data show a mild preference for a closed Universe (\text{$\Ok<0$}), hinting at a small but notable tension with the flat \lcdm\ prediction.

\begin{figure}[!ht]
	\centering
	\includegraphics[width=.85\columnwidth]{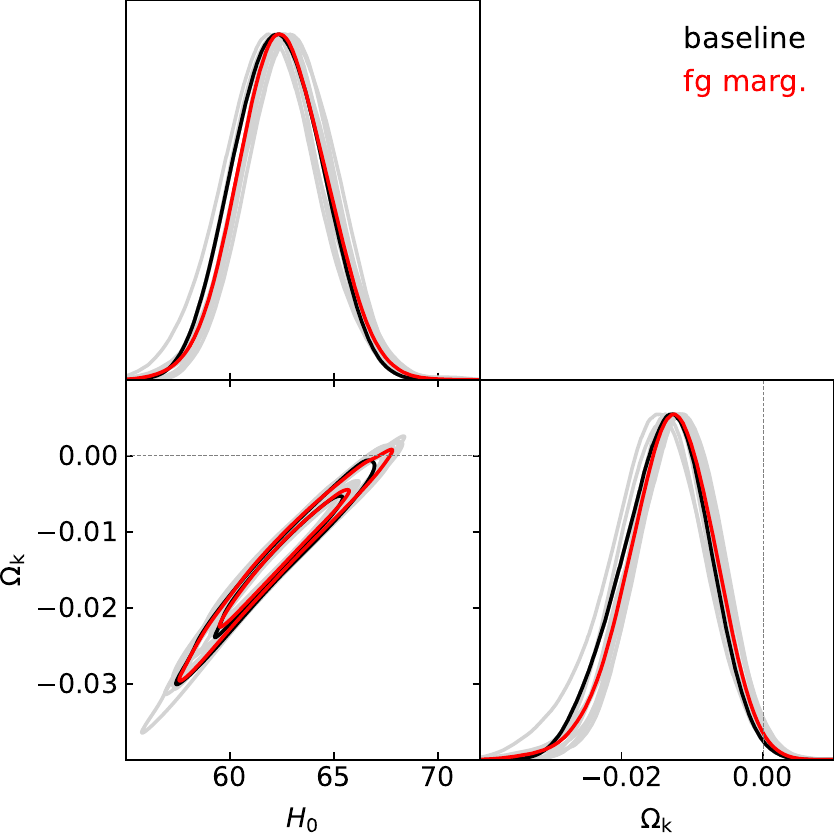}
	\caption{Posterior distributions for \Ok\ combining \PLK, \ACT, and \SPT datasets, with varying foreground templates (\textit{in grey}), compared to the posteriors after marginalisation over the foreground models (\textit{in red}).}
	\label{fig:Ok_marg}
\end{figure}

Figure~\ref{fig:Ok_marg} shows the posterior for $\Ok$ together with its geometric degeneracy with $H_0$. As discussed in \citetalias{tristram:2024}, with \PLK\ PR4, the tail of the 2D posterior in the $H_0$–$\Ok$ plane at low $H_0$ and negative $\Ok$ is significantly less preferred. Indeed, using the combined CMB datasets, averaging over foreground models,  
$$\Ok =  -0.0132\,^{+0.0066}_{-0.0056}\text{(stat)} \pm 0.0009\text{(fg)} \quad \text{(\PLK+\ACT+\SPT)}.$$

The central value is fully compatible with that found in \PLK\ PR4 analyses and lower than the value inferred with PR3. while the uncertainties are about a factor of two lower than in PR4 and three times lower than in PR3. As for PR4, and unlike PR3, we do not find any significant evidence for non-zero spatial curvature.

\section{Conclusions}
\label{sec:conclusion}

In this work, we constructed a CMB likelihood that combines the latest data from \PLK, \ACT, and \SPT\ within a single, coherent framework. We took particular care to implement a consistent treatment of both foreground modelling and instrumental systematics, to ensure that the different datasets can be robustly analysed together. 
Compared to a simple multiplication of the individual likelihoods, our approach explicitly enforces consistency in the modelling assumptions and parametrisations across experiments. This is essential for disentangling true cosmological signals from experiment-specific systematics or foreground treatments, and it allows us to quantify uncertainties in a uniform way. Crucially, we can also then marginalise over the choice of foreground templates, ensuring that the resulting cosmological constraints reflect not only statistical precision but also the systematic uncertainty associated with different modelling assumptions.

We first examined the impact of this joint likelihood on the \lcdm\ parameters. The combined constraints are consistent with those obtained from the individual experiments, and the parameter means show negligible dependence on the details of the foreground modelling. Marginalising over the foreground templates increases the uncertainties by less than 5\%, except for $n_{\rm s}$ (9\%). Compared to \PLK\ alone, the joint analysis improves the precision of the parameters by up to $20\%$, demonstrating the complementarity of the datasets and the benefit of a coherent and combined analysis.

Turning to foreground parameter constraints, we observe a stronger dependence on modelling choices. In particular, the reconstruction of the CIB amplitude, the kSZ signal, and the tSZ$\times$CIB correlation show significant sensitivity to the adopted templates, while the tSZ is relatively more stable. 
The uncertainties on $A_{\rm CIB}$ (and, to a lesser extent, on $A_{\rm kSZ}$ and $\xi_{\rm tSZ\times CIB}$) are dominated by variations in the foreground modelling rather than statistical noise.
The treatment of point-source and dust residuals emerge as the most important factors influencing the inferred amplitudes of extragalactic foregrounds. This underlines the necessity of a unified multi-dataset framework to achieve reliable astrophysical constraints.

We also investigated the impact of the joint likelihood on several simple extensions to \lcdm. For the lensing amplitude $\Alens$, the sum of neutrino masses $\Mnu$, the curvature parameter $\Omega_k$, and the effective number of relativistic species $\Neff$, the combination of datasets provides substantial gains in precision, with no evidence for departures from \lcdm\ expectations. 
Several previously reported tensions -- such as hints of non-zero curvature, or values of $\Alens>1$ -- are significantly reduced or even resolved when using PR4 data and the combined likelihood. We further find that uncertainties in foreground modelling have a non-negligible effect, increasing the error bars by up to 35\%.

Overall, this work demonstrates that a coherent combination of \PLK, \ACT, and \SPT\ data is feasible and robust. While cosmological constraints remain stable against foreground modelling, the treatment of foregrounds emerges as an important source of uncertainty for \lcdm\ extensions. The framework developed here provides a solid basis for future analyses, particularly those aiming to incorporate more sophisticated physical models, such as halo-based descriptions of extragalactic components \citep[as done for instance in][]{Douspis:2022,Gorce:2022,Bolliet:2025}, that can propagate astrophysical uncertainties when fitting CMB data in a controlled manner. Such developments will be crucial for the next generation of high-sensitivity surveys, including the Simons Observatory and the South Pole Observatory, where the precision of cosmological inference will increasingly depend on the accuracy of foreground modelling. In this context, the combination with \PLK data will continue to play a central role, both as a calibration anchor and as a unique source of large-scale information.

\begin{acknowledgements}
The framework used in the analysis is available on github (\href{https://github.com/mtristram/hillik}{github.com/mtristram/hillik}).
We acknowledge support from the French Agence Nationale de la Recherche (ANR), under grant ANR-22-CE31-0010 (project BATMAN). We also thank the developers of the publicly available software used in this work: \texttt{Cobaya}~\citep{torrado:2021}, \texttt{CAMB}~\citep{Lewis:2000}, \texttt{CosmoPower}~\citep{Mancini:2022}, and \texttt{GetDist}~\citep{Lewis:2025}.
We gratefully acknowledge support from the CNRS/IN2P3 Computing Center for providing computing and data-processing resources needed for this work. 
\end{acknowledgements}

\bibliographystyle{aa}
\bibliography{aa58015-25}

\newpage
\onecolumn

\begin{appendix}
    
\section{Correlations}

\begin{figure}[!htbp]
	\centering
	\includegraphics[width=\columnwidth]{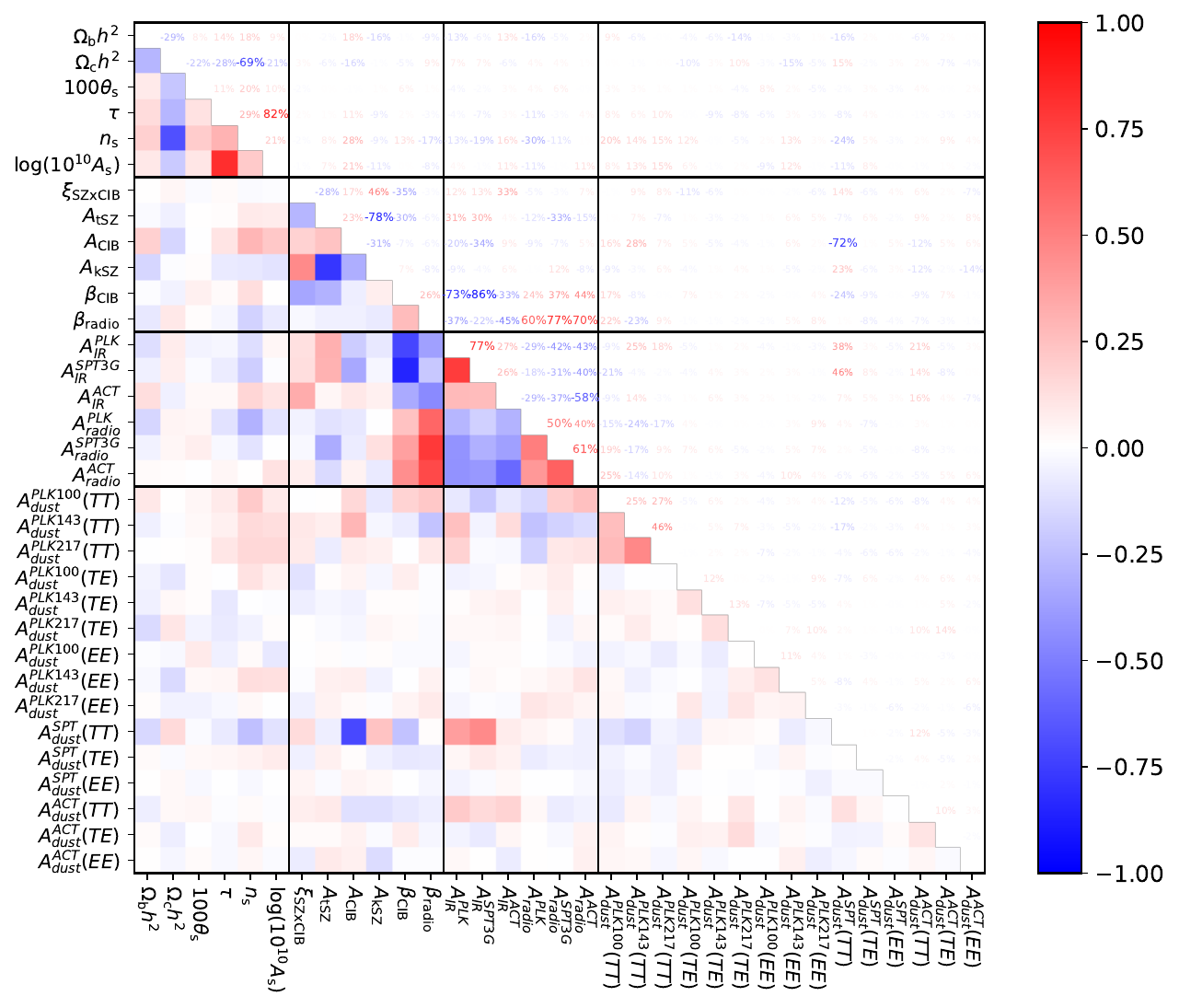}
	\caption{Correlation matrix for the fitted parameters of the combined likelihood TTTEEE. The first block corresponds to cosmological parameters from the \lcdm\ model, the second block gathers the foreground parameters common to all datasets, the third and fourth blocks show, for each dataset, the amplitude of the point sources and the dust amplitudes respectively.}
	\label{fig:corr}
\end{figure}

Figure~\ref{fig:corr} shows the correlation matrix for the cosmological and foreground parameters. The upper triangular half of the matrix displays the correlation coefficients, while the lower triangular half visualises them using the color scale.

\clearpage
\section{Foreground parameters}

\begin{table*}[!ht]
    \renewcommand{\arraystretch}{1.1}
	\begin{center}
    \caption{Foreground parameters.}
	\begin{tabular}{lccccl}
	\hline
	\hline
	Likelihood     & Parameter     &  Prior  &  Posterior  & Spectrum & Description  \\
\hline Common
    & $A_\mathrm{CIB}$ & $[0,50]$ & $ 4.32\pm 0.73$ & $TT$ & Clustered CIB amplitude\\
    & $A_\mathrm{tSZ}$ & $[0,50]$ & $ 2.98\pm 0.25$ & $TT$ & tSZ amplitude\\
    & $A_\mathrm{kSZ}$ & $[0,50]$ & $ 1.20\pm 0.87$ & $TT$ & kSZ amplitude\\
    & $\xi_\mathrm{SZxCIB}$ & $[-1,1]$ & $0.044\pm0.064$ & $TT$ & SZ$\times$CIB amplitude\\
    & $\beta_\mathrm{CIB}$ & $[1,3]$ & $ 1.79\pm 0.19$ & $TT$ & CIB spectral index\\
    & $\beta_\mathrm{radio}$ & $[-1.5,0.0]$ & $-0.88\pm 0.05$ & $TT$ & Radio sources spectral index\\
    & $\beta_{\rm d}^{TT}$ & $\mathcal{N}(1.51, 0.01)$ & $1.510\pm0.009$ & $TT$ & Dust $TT$ spectral index\\
    & $\beta_{\rm d}^{TE}$ & $\mathcal{N}(1.59, 0.02)$ & $1.588\pm0.021$ & $TE$ & Dust $TE$ spectral index\\
    & $\beta_{\rm d}^{EE}$ & $\mathcal{N}(1.59, 0.02)$ & $1.612\pm0.020$ & $EE$ & Dust $EE$ spectral index\\
    & $\alpha_{\rm d}^{TT}$ & $[-3,-2]$ & $-2.62\pm 0.07$ & $TT$ & Power-law index of the $TT$ dust spectrum\\
    & $\alpha_{\rm d}^{TE}$ & $-2.4$ & $-2.4$ & $TE$ & Power-law index of the $TE$ dust spectrum\\
    & $\alpha_{\rm d}^{EE}$ & $-2.4$ & $-2.4$ & $EE$ & Power-law index of the $EE$ dust spectrum\\
\hline $Planck$
    & $A_\mathrm{dust}^\mathrm{PLK100}(TT)$ & $[0,100]$ & $ 24.2\pm  9.0$ & $TT$ & Dust amplitude in $TT$ in 100\GHz\ maps\\
    & $A_\mathrm{dust}^\mathrm{PLK143}(TT)$ & $[0,100]$ & $ 21.0\pm  1.8$ & $TT$ & Dust amplitude in $TT$ in 143\GHz\ maps\\
    & $A_\mathrm{dust}^\mathrm{PLK217}(TT)$ & $[0,100]$ & $ 9.71\pm 0.72$ & $TT$ & Dust amplitude in $TT$ in 217\GHz\ maps\\
    & $A_\mathrm{dust}^\mathrm{PLK100}(TE)$ & $[0,10]$ & $1.477\pm0.320$ & $TE$ & Dust amplitude in $TE$ in 100\GHz\ maps\\
    & $A_\mathrm{dust}^\mathrm{PLK143}(TE)$ & $[0,10]$ & $0.792\pm0.151$ & $TE$ & Dust amplitude in $TE$ in 143\GHz\ maps\\
    & $A_\mathrm{dust}^\mathrm{PLK217}(TE)$ & $[0,10]$ & $0.434\pm0.044$ & $TE$ & Dust amplitude in $TE$ in 217\GHz\ maps\\
    & $A_\mathrm{dust}^\mathrm{PLK100}(EE)$ & $[0,10]$ & $0.628\pm0.201$ & $EE$ & Dust amplitude in $EE$ in 100\GHz\ maps\\
    & $A_\mathrm{dust}^\mathrm{PLK143}(EE)$ & $[0,10]$ & $0.293\pm0.012$ & $EE$ & Dust amplitude in $EE$ in 143\GHz\ maps\\
    & $A_\mathrm{dust}^\mathrm{PLK217}(EE)$ & $[0,10]$ & $0.174\pm0.017$ & $EE$ & Dust amplitude in $EE$ in 217\GHz\ maps\\
    & $A_\mathrm{radio}^\mathrm{PLK}$ & $[0,150]$ & $58.06\pm11.41$ & $TT$ & Unresolved radio sources\\
    & $A_\mathrm{IR}^\mathrm{PLK}$ & $[0,100]$ & $ 9.71\pm11.60$ & $TT$ & Unresolved infrared sources\\
\hline ACT
    & $A_\mathrm{dust}^\mathrm{ACT}(TT)$ & $\mathcal{N}(7.95, 0.32)$ & $ 7.81\pm 0.53$ & $TT$ & Dust amplitude in $TT$\\
    & $A_\mathrm{dust}^\mathrm{ACT}(TE)$ & $\mathcal{N}(0.423, 0.030)$ & $0.413\pm0.051$ & $TE$ & Dust amplitude in $TE$\\
    & $A_\mathrm{dust}^\mathrm{ACT}(EE)$ & $\mathcal{N}(0.168, 0.017)$ & $0.185\pm0.080$ & $EE$ & Dust amplitude in $EE$\\
    & $A_\mathrm{radio}^\mathrm{ACT}$ & $[0,20]$ & $ 3.01\pm 2.21$ & $TT$ & Unresolved radio sources in $TT$\\
    & $A_\mathrm{IR}^\mathrm{ACT}$ & $[0,20]$ & $ 7.45\pm 1.52$ & $TT$ & Unresolved infrared sources\\
\hline SPT
    & $A_\mathrm{dust}^\mathrm{SPT}(TT)$ & $\mathcal{N}(1.88, 0.960)$ & $ 2.30\pm 1.03$ & $TT$ & Dust amplitude in $TT$\\
    & $A_\mathrm{dust}^\mathrm{SPT}(TE)$ & $\mathcal{N}(0.12, 0.051)$ & $0.105\pm0.033$ & $TE$ & Dust amplitude in $TE$\\
    & $A_\mathrm{dust}^\mathrm{SPT}(EE)$ & $\mathcal{N}(0.05, 0.022)$ & $0.057\pm0.013$ & $EE$ & Dust amplitude in $EE$\\
    & $A_\mathrm{radio}^\mathrm{SPT3G}$ & $[0,10]$ & $ 0.68\pm 0.06$ & $TT$ & Unresolved radio sources in $TT$\\
    & $A_\mathrm{IR}^\mathrm{SPT3G}$ & $[0,10]$ & $ 7.33\pm 0.64$ & $TT$ & Unresolved infrared sources\\
    \hline
    \end{tabular}
    \tablefoot{For each parameter we provide the prior and constraint after marginalisation over the foreground templates (mean and standard deviation). In total, this amounts to 31 parameters: 10 parameters common to all datasets, 11 parameters specific to \PLK, 5 to \ACT, and 5 to \SPT. The datasets share the same set of foreground parameters, except for the dust and point sources amplitudes. The dust amplitudes depend on the sky fraction: \PLK\ includes three amplitudes, while \ACT\ and \SPT\ each have only one, corresponding to their single mask survey.}
    \label{tab:fgpar}
    \end{center}
\end{table*}

\section{Nuisance parameters}

\begin{table*}[!ht]
    \renewcommand{\arraystretch}{1.05}
	\begin{center}
    \caption{Nuisance parameters for instrumental effects.}
	\begin{tabular}{lcccl}
	\hline
	\hline
	Likelihood     & Parameter     &  Prior & Posterior  & Description \\
\hline $Planck$
    & $A_\mathrm{planck}$ & $\mathcal{N}(1.00,0.0025)$ & $1.001\pm0.002$  & Absolute calibration uncertainty\\
    & $\mathrm{c}^\mathrm{PLK}_{100A}$ & $\mathcal{N}(1.00,0.01)$ & $1.0007\pm0.0074$  & Intercalibration map 100A\\
    & $\mathrm{c}^\mathrm{PLK}_{100B}$ & $\mathcal{N}(1.00,0.01)$ & $0.9961\pm0.0076$  & Intercalibration map 100B\\
    & $\mathrm{c}^\mathrm{PLK}_{143A}$ & $1.00 \text{(fixed)}$ & $1.00$  & Calibration reference map 143A\\
    & $\mathrm{c}^\mathrm{PLK}_{143B}$ & $\mathcal{N}(1.00,0.01)$ & $1.0002\pm0.0072$  & Intercalibration map 143B\\
    & $\mathrm{c}^\mathrm{PLK}_{217A}$ & $\mathcal{N}(1.00,0.01)$ & $1.0026\pm0.0080$  & Intercalibration map 217A\\
    & $\mathrm{c}^\mathrm{PLK}_{217B}$ & $\mathcal{N}(1.00,0.01)$ & $1.0001\pm0.0079$  & Intercalibration map 217B\\
    & $\rho^\mathrm{PLK}_{100A}$ & $[0.8,1.2]$ & $1.007\pm0.032$  & Polarisation efficiency map 100A\\
    & $\rho^\mathrm{PLK}_{100B}$ & $[0.8,1.2]$ & $1.013\pm0.030$  & Polarisation efficiency map 100B\\
    & $\rho^\mathrm{PLK}_{143A}$ & $[0.8,1.2]$ & $0.962\pm0.025$  & Polarisation efficiency map 143A\\
    & $\rho^\mathrm{PLK}_{143B}$ & $[0.8,1.2]$ & $1.022\pm0.026$  & Polarisation efficiency map 143B\\
    & $\rho^\mathrm{PLK}_{217A}$ & $[0.8,1.2]$ & $1.035\pm0.041$  & Polarisation efficiency map 217A\\
    & $\rho^\mathrm{PLK}_{217B}$ & $[0.8,1.2]$ & $1.021\pm0.041$  & Polarisation efficiency map 217B\\
\hline ACT
    & $\mathrm{A}^\mathrm{ACT}$ & $[0.7,1.3]$ & $0.9868\pm0.0052$  & \ACT\ calibration\\
    & $\mathrm{c}^\mathrm{ACT}_{090pa5}$ & $\mathcal{N}(1.00,0.0016)$ & $0.9981\pm0.0013$  & Intercalibration pa5 90\GHz\\
    & $\mathrm{c}^\mathrm{ACT}_{090pa6}$ & $\mathcal{N}(1.00,0.0018)$ & $0.9990\pm0.0014$  & Intercalibration pa6 90\GHz\\
    & $\mathrm{c}^\mathrm{ACT}_{150pa5}$ & $1.00 \text{(fixed)}$ & $1.00$  & Calibration reference (pa5 150\GHz)\\
    & $\mathrm{c}^\mathrm{ACT}_{150pa6}$ & $\mathcal{N}(1.00,0.0024)$ & $1.0038\pm0.0016$  & Intercalibration pa6 150\GHz\\
    & $\mathrm{c}^\mathrm{ACT}_{220pa4}$ & $\mathcal{N}(1.00,0.013)$ & $0.9908\pm0.0079$  & Intercalibration pa4 220\GHz\\
    & $\rho^\mathrm{ACT}_{090pa5}$ & $[0.8,1.2]$ & $1.0015\pm0.0042$  & Polarisation efficiency pa5 90\GHz\\
    & $\rho^\mathrm{ACT}_{090pa6}$ & $[0.8,1.2]$ & $1.0109\pm0.0044$  & Polarisation efficiency pa6 90\GHz\\
    & $\rho^\mathrm{ACT}_{150pa5}$ & $[0.8,1.2]$ & $1.0067\pm0.0043$  & Polarisation efficiency pa5 150\GHz\\
    & $\rho^\mathrm{ACT}_{150pa6}$ & $[0.8,1.2]$ & $1.0085\pm0.0048$  & Polarisation efficiency pa6 150\GHz\\
    & $\Delta^\mathrm{ACT}_{090pa5}$ & $\mathcal{N}(0.0,1.0)$ & $0.46\pm0.63$  & Bandpass uncertainty pa5 90\GHz\\
    & $\Delta^\mathrm{ACT}_{090pa6}$ & $\mathcal{N}(0.0,1.2)$ & $0.92\pm0.79$  & Bandpass uncertainty pa6 90\GHz\\
    & $\Delta^\mathrm{ACT}_{150pa5}$ & $\mathcal{N}(0.0,1,3)$ & $-1.64\pm0.83$  & Bandpass uncertainty pa5 150\GHz\\
    & $\Delta^\mathrm{ACT}_{150pa6}$ & $\mathcal{N}(0.0,1.1)$ & $-1.00\pm0.76$  & Bandpass uncertainty pa6 150\GHz\\
    & $\Delta^\mathrm{ACT}_{220pa4}$ & $\mathcal{N}(0.0,3.6)$ & $0.06\pm1.68$  & Bandpass uncertainty pa4 220\GHz\\
\hline SPT
    & $\mathrm{A}^\mathrm{SPT}$ & $[0.7,1.3]$ & $1.0013\pm0.0050$  & \SPT\ calibration\\
    & $\mathrm{c}^\mathrm{SPT}_{90}$ & $\mathcal{N}(1.00,0.0003)$ & $1.0004\pm0.0004$  & Intercalibration 90\GHz\\
    & $\mathrm{c}^\mathrm{SPT}_{150}$ & $1.00 \text{(fixed)}$ & $1.00$  & Calibration reference 150\GHz\\
    & $\mathrm{c}^\mathrm{SPT}_{220}$ & $\mathcal{N}(1.00,0.001)$ & $1.0092\pm0.0011$  & Intercalibration 220\GHz\\
    & $\rho^\mathrm{SPT}_{90}$ & $[0.8,1.2]$ & $1.0014\pm0.0039$  & Polarisation efficiency 90\GHz\\
    & $\rho^\mathrm{SPT}_{150}$ & $[0.8,1.2]$ & $1.0030\pm0.0039$  & Polarisation efficiency 150\GHz\\
    & $\rho^\mathrm{SPT}_{220}$ & $[0.8,1.2]$ & $0.9981\pm0.0049$  & Polarisation efficiency 220\GHz\\
    & $\beta^\mathrm{SPT}_{1}$ & $\mathcal{N}(0.0,1.0)$ & $-1.15\pm0.94$  & Main beam eigenmodes\\
    & $\beta^\mathrm{SPT}_{2}$ & $\mathcal{N}(0.0,1.0)$ & $-0.50\pm0.88$  & Main beam eigenmodes\\
    & $\beta^\mathrm{SPT}_{3}$ & $\mathcal{N}(0.0,1.0)$ & $0.03\pm0.92$  & Main beam eigenmodes\\
    & $\beta^\mathrm{SPT}_{4}$ & $\mathcal{N}(0.0,1.0)$ & $-1.85\pm0.63$  & Main beam eigenmodes\\
    & $\beta^\mathrm{SPT}_{5}$ & $\mathcal{N}(0.0,1.0)$ & $0.24\pm0.92$  & Main beam eigenmodes\\
    & $\beta^\mathrm{SPT}_{6}$ & $\mathcal{N}(0.0,1.0)$ & $-1.40\pm0.90$  & Main beam eigenmodes\\
    & $\beta^\mathrm{SPT}_{7}$ & $\mathcal{N}(0.0,1.0)$ & $0.15\pm0.72$  & Main beam eigenmodes\\
    & $\beta^\mathrm{SPT}_{8}$ & $\mathcal{N}(0.0,1.0)$ & $-0.20\pm0.98$  & Main beam eigenmodes\\
    & $\beta^\mathrm{SPT}_{9}$ & $\mathcal{N}(0.0,1.0)$ & $0.42\pm0.97$  & Main beam eigenmodes\\
    & $\beta^\mathrm{SPT}_{90}$ & $[0.0,1.0]$ & $0.555\pm0.079$  & Beam sidelobe polarisation fraction 90\GHz\\
    & $\beta^\mathrm{SPT}_{150}$ & $[0.0,1.0]$ & $0.716\pm0.100$  & Beam sidelobe polarisation fraction 150\GHz\\
    & $\beta^\mathrm{SPT}_{220}$ & $[0.0,1.0]$ & $0.681\pm0.124$  & Beam sidelobe polarisation fraction 220\GHz\\
    & $\epsilon^\mathrm{SPT}_{90}$ & $\mathcal{N}(-0.0065,0.0011)$ & $-0.0073\pm0.0008$  & Temperature-to-polarisation leakage 90\GHz\\
    & $\epsilon^\mathrm{SPT}_{150}$ & $\mathcal{N}(-0.012,0.0021)$ & $-0.0154\pm0.0014$  & Temperature-to-polarisation leakage 150\GHz\\
    & $\epsilon^\mathrm{SPT}_{220}$ & $\mathcal{N}(-0.023,0.0066)$ & $-0.0291\pm0.0036$  & Temperature-to-polarisation leakage 220\GHz\\
    & $\kappa^\mathrm{SPT}$ & $\mathcal{N}(0.00000,0.00045)$ & $-0.00008\pm0.00029$  & Super-sample lensing\\
    \hline
    \end{tabular}
    \tablefoot{For each parameter we provide the prior and constraint after marginalisation over the foreground templates (mean and standard deviation). This results in 48 parameters: 12~parameters for \PLK, 14~parameters for \ACT, and 22~parameters for \SPT.}
    \label{tab:nuipar}
    \end{center}
\end{table*}

Table~\ref{tab:nuipar} presents all nuisance parameters together with their constraint for the combination of \PLK, \ACT, and \SPT\ data after marginalising over the foreground templates.

While \PLK\ is calibrated using the orbital dipole (through the satellite velocity and the CMB monopole), ground-based experiments usually rely on measurements of planet fluxes to derive their map calibration. In a second step, their spectra can be re-calibrated by cross-correlating with \PLK\ maps.
As mentioned in Sect.~\ref{sec:nui}, we allow the datasets to recalibrate on \PLK. This is done by introducing calibration factors applied to \SPT, and \ACT\ independently.

We find a good calibration consistency between the datasets, with $c_\mathrm{\SPT} = 1.0011 \pm 0.0043$ and $c_\mathrm{\ACT}  = 0.9868 \pm 0.0049$, although, for \ACT, we find a deviation of -1.3\%, significantly greater than the absolute calibration accuracy of 0.3\% reported by \citetalias{Louis:2025}. In contrast, for \SPT\ our analysis yields an overall recalibration of 0.11\%, consistent with the absolute calibration uncertainty of 0.36\% reported by \citetalias{Camphuis:2025}.

All polar efficiencies are found to be compatible with unity at better than $1\sigma$. The uncertainties depend on the dataset with $\sim$3\%, 2.5\%, and 4\% for \PLK\ at 100, 143 and 217\GHz\ respectively, and $\sim$0.4\% for \SPT\ or \ACT.

\clearpage
\section{Extensions}
\label{ann:ext}

\begin{figure*}[!ht]
    \centering
	\includegraphics[width=.3\textwidth,height=120pt]{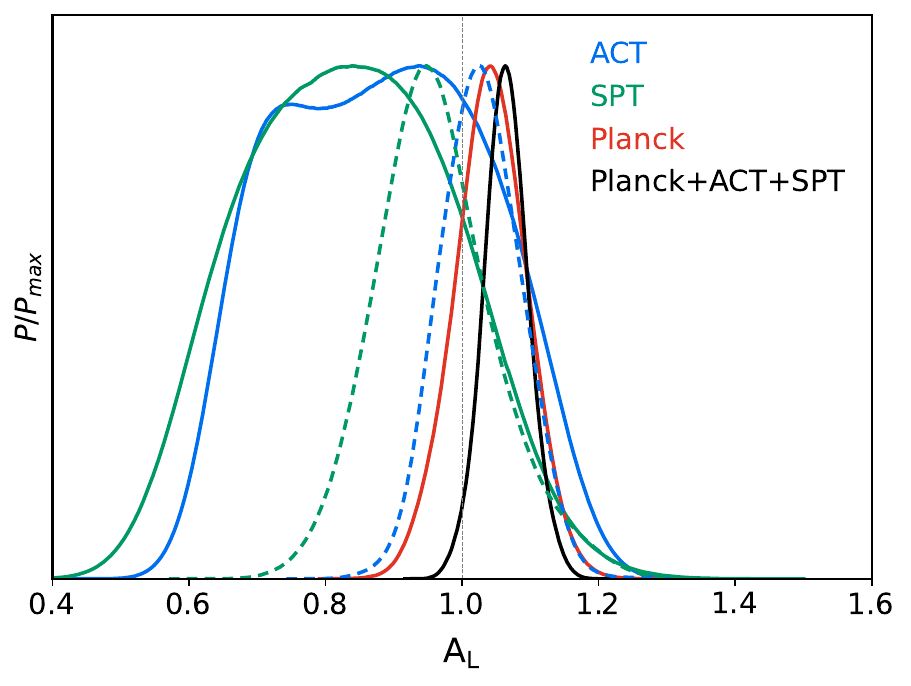}
    \hspace{10pt}
	\includegraphics[width=.3\textwidth,height=120pt]{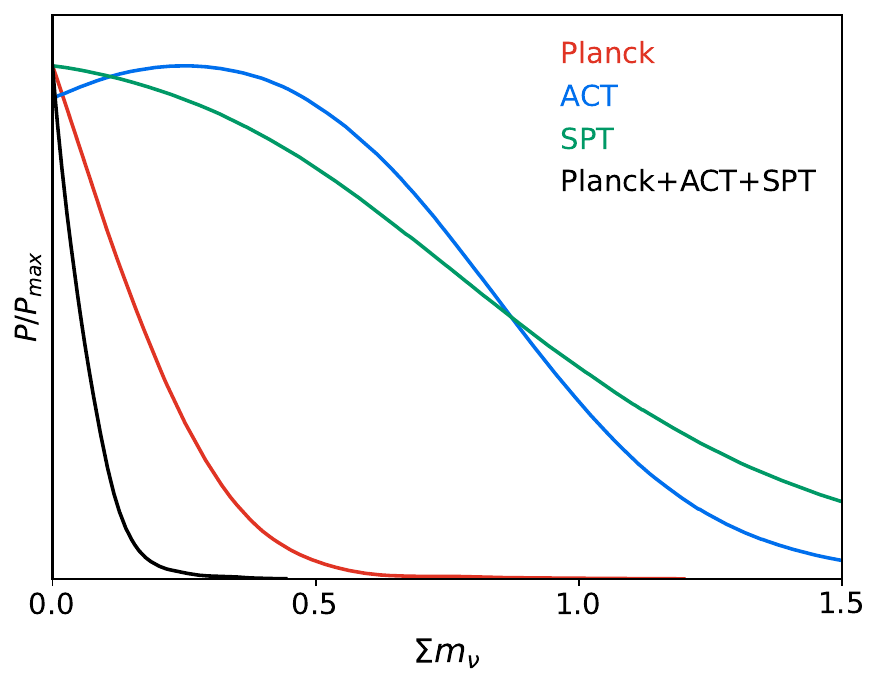}
    \hspace{10pt}
	\includegraphics[width=.3\textwidth,height=120pt]{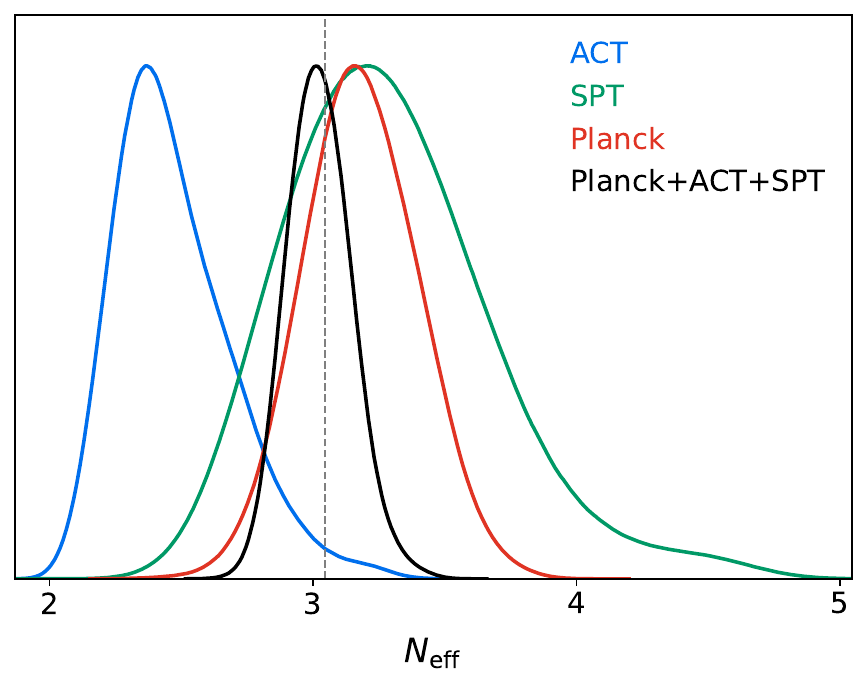}
	\caption{Posterior distributions for \Alens\ (left), \Mnu\ (center), and \Neff\ (right) using \PLK, \ACT, or \SPT, and their combination. The dashed lines on the left plot represent the posterior distribution with a prior on the reionisation optical depth, $\tau=\mathcal{N}(0.057,0.006)$.}
	\label{fig:ext}
\end{figure*}

\subsection{Amplitude of the gravitational lensing, \texorpdfstring{\Alens}{Alens}}

We vary $\Alens$ for the three CMB datasets and derived the corresponding marginal posteriors, shown in Fig.~\ref{fig:ext}. For all datasets, the recovered values of $\Alens$ are consistent with unity, indicating that the smoothing of the acoustic peaks in the CMB spectra is compatible with \lcdm\ expectations:
\begin{align}
    \Alens &= 1.042 \pm 0.052 \quad \text{(\PLK)}  \nonumber\\
    \Alens &= 0.881 \pm 0.175 \quad \text{(\ACT)} \nonumber\\
    \Alens &= 0.833 \pm 0.174 \quad \text{(\SPT)}  \nonumber
\end{align}
For \PLK, the result is fully consistent with \citetalias{tristram:2024}, which report $\Alens = 1.039 \pm 0.052$. The uncertainties for \ACT\ and \SPT\ are larger, primarily due to the degeneracy with $A_{\rm s}$ and $\tau$, which remains unresolved without additional constraints on the reionisation optical depth. When a prior on $\tau$ is applied, $\tau=\mathcal{N}(0.057,0.006)$, we obtain $\Alens = 1.027 \pm 0.058$ for \ACT\ and $\Alens = 0.961 \pm 0.089$ for \SPT, values that are closer to those reported in \citetalias{Louis:2025} and \citetalias{Camphuis:2025}, respectively.

\subsection{Sum of the neutrino masses, \texorpdfstring{\Mnu}{Mnu}}
Figure~\ref{fig:ext} shows the posterior distributions for the sum of the neutrino masses, $\Mnu$. No evidence for a non-zero neutrino mass is found, and we report the following upper limits:
\begin{align}
    \Mnu &< 0.402\,\text{eV} \quad \text{(95\% CL, \PLK)} \nonumber \\
    \Mnu &< 1.120\,\text{eV} \quad \text{(95\% CL, \ACT)} \nonumber\\
    \Mnu &< 1.447\,\text{eV} \quad \text{(95\% CL, \SPT)} \nonumber
\end{align}

\subsection{Effective number of relativistic species, \texorpdfstring{\Neff}{Neff}}
Figure~\ref{fig:ext} shows the posteriors for \PLK, \ACT, \SPT, and their combination when we consider the \Neff\ extension.
The individual datasets yield:
\begin{align}
    \Neff &= 3.20^{+0.21}_{-0.23} \quad \text{(\PLK)}\nonumber \\
    \Neff &= 2.45^{+0.17}_{-0.25} \quad \text{(\ACT)} \nonumber\\
    \Neff &= 3.23^{+0.32}_{-0.38} \quad \text{(\SPT)} \nonumber
\end{align}
These results are consistent with those obtained in the original analyses, which quote $\Neff = 2.60^{+0.21}_{-0.29}$ for \ACT\ \citep{Calabrese:2025}, $\Neff = 3.17^{+0.29}_{-0.33}$ for \SPT\ \citepalias{Camphuis:2025}, and $\Neff = 3.08 \pm 0.17$ for \PLK\ PR4 \citepalias{tristram:2024}. In particular, we reproduce the tendency of \ACT\ to shift $\Neff$ toward low values.

\subsection{Spatial curvature, \texorpdfstring{\Ok}{Ok}}

We find that CMB data are compatible with a flat Universe, with $\Ok$ consistent with zero for all datasets:
\begin{align}
    \Ok &= -0.015^{+0.011}_{-0.009}             \quad \text{(\PLK)} \nonumber \\
    \Ok &= -0.006^{+0.020}_{-0.016}             \quad \text{(ACT)}  \nonumber\\
    \Ok &= \phantom{+}0.003^{+0.019}_{-0.014}   \quad \text{(SPT)}  \nonumber
\end{align}
These results are compatible with those of \citetalias{Louis:2025} and \citetalias{Camphuis:2025}, although with larger error bars due to the absence of a $\tau$-prior. For \PLK, our constraint is comparable to the \hillipop\ PR4 result \citepalias[$\Ok = -0.012 \pm 0.010$,][]{tristram:2024} and should be contrasted with $\Ok = -0.025_{-0.010}^{+0.013}$ from \camspec\ PR4 \citep{rosenberg:2022} and $\Ok = -0.044_{-0.015}^{+0.018}$ from \plik\ PR3 \citep{planck2016-l06}.

\begin{figure}[!ht]
	\centering
	\includegraphics[width=.4\columnwidth]{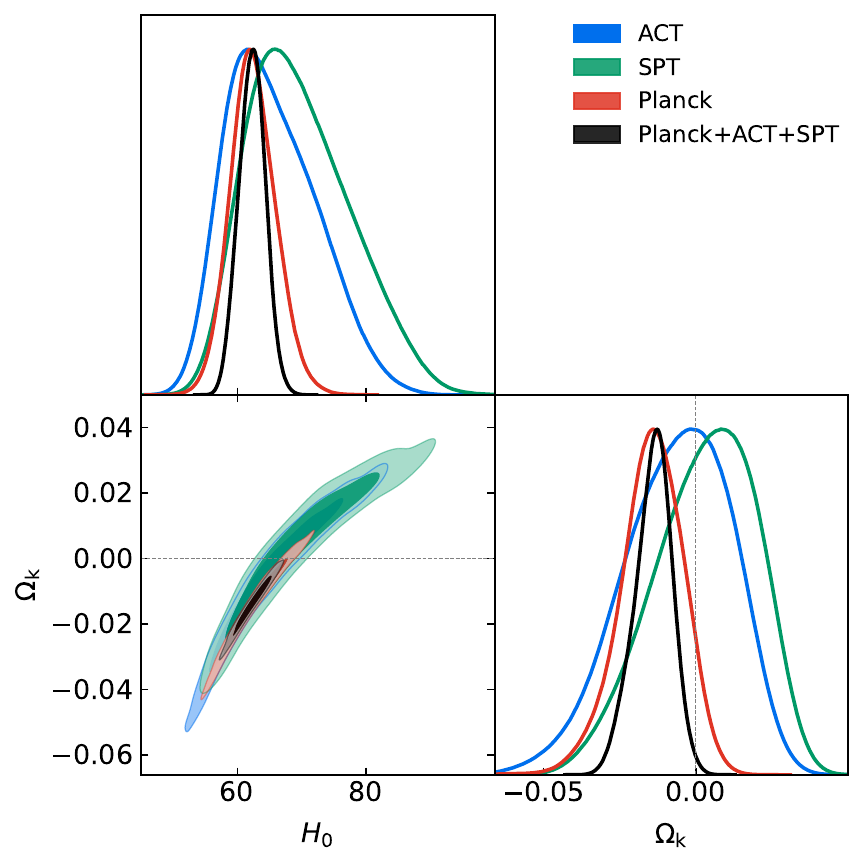}
	\caption{Posterior distributions for \Ok\ using \PLK, \ACT, or \SPT, and their combination.}
	\label{fig:Ok}
\end{figure}

Figure~\ref{fig:Ok} shows the posterior for $\Ok$ together with its geometric degeneracy with $H_0$. As discussed in \citetalias{tristram:2024}, with \PLK\ PR4, the tail of the 2-d posterior in the $H_0$–$\Ok$ plane at low $H_0$ and negative $\Ok$ is less preferred compared to \PLK\ PR3. 

\section{Goodness-of-fit metric}
This appendix reports the $\Delta\chi^2 = \chi^2_\mathrm{variation} - \chi^2_\mathrm{baseline}$ values (baseline defined in Sect.~\ref{sec:model}) for all foreground models, reported per likelihood and for their combination. The overall $\chi^2$ variation is small (below $26.4$, for 7403 data points), indicating that no model offers a significantly better or worse fit.

\begin{table*}[!ht]
    \renewcommand{\arraystretch}{1.}
	\begin{center}
    \caption{Goodness of fit.}
	\begin{tabular}{ll|cccc}
	\hline
	\hline
	Foreground     & Model     &  \multicolumn{4}{c}{$\Delta\chi^2$} \\
	     &      &  PLK+ACT+SPT & PLK & ACT & SPT \\
	\hline
CIB & Addison et al. 2012 & $-9.2$ &  $0.3$ & $-3.0$ & $-6.5$ \\
CIB & Lenz et al. 2019 &  $5.3$ &  $5.9$ & $-5.4$ &  4.5 \\
CIB & Viero et al. 2013 & $-3.9$ &  $2.4$ & $-0.3$ & $-1.9$ \\
CIB & Mak et al. 2017 & $-9.8$ & $-2.0$ & $-2.1$ & $-4.3$ \smallskip\\
tSZ & Battaglia et al. 2012 &  $1.7$ & $-1.1$ & $-0.7$ &  $4.3$ \\
tSZ & Omori 2024 & $-0.3$ & $-0.1$ & $-5.5$ &  $1.6$ \\
tSZ & Efstathiou\&Migliaccio 2012 & $-7.5$ & $-0.9$ &  $1.3$ &  $4.4$ \\
tSZ & Shaw et al. 2010 & $-2.9$ &  $1.3$ &  $0.4$ &  $1.4$ \smallskip\\
kSZ & Battaglia et al. 2010 & $-3.9$ & $-0.6$ &  $2.2$ &  $3.6$ \\
kSZ & Trac et al. 2011 &  $2.1$ &  $0.0$ &  $1.4$ &  $0.8$ \\
kSZ & Omori 2024 & $-2.7$ & $-3.3$ & $-1.5$ &  $4.6$ \\
kSZ & Gorce et al. 2020 & $-1.8$ & $-0.2$ &  $1.4$ &  $2.0$ \smallskip\\
tSZxCIB & Zahn et al. 2012 & $-0.3$ & $-0.9$ & $-0.7$ &  $4.7$ \\
tSZxCIB & $\sqrt{C_\ell^{CIB} * C_\ell^{tSZ}}$ &  $0.6$ & $-0.1$ & $-1.0$ & $-1.7$ \\
tSZxCIB & Maniyar et al. 2021 & $-2.7$ & $-0.2$ &  $1.8$ & $-3.2$ \\
tSZxCIB & Addison et al. 2012 & $-2.2$ & $-0.5$ & $-2.4$ &  $3.7$ \smallskip\\
PS & $\beta_\mathrm{DSFG} \neq \beta_\mathrm{CIB}$ & $-1.1$ & $-3.2$ &  $1.0$ &  $4.8$ \\
Dust & ($\beta_\mathrm{d}$ prior, $A_\mathrm{dust}$ free) & $-5.1$ & $ 1.2$ &  $2.6$ & $-2.7$ \\
Dust & ($\beta_\mathrm{d}$ free, $A_\mathrm{dust}$ prior) & $-21.1$ & $-22.9$ & $-1.7$ & $-0.3$ \\
    \hline
    \end{tabular}
    \label{tab:chi2}
   \tablefoot{$\Delta\chi^2 = \chi^2_\mathrm{variation} - \chi^2_\mathrm{baseline}$ for \lcdm. The total data vector size is 7403 (PLK=4872, ACT=1139, SPT=1392).}
    \end{center}
\end{table*}

\end{appendix}

\end{document}